\documentclass[sigconf]{acmart}

\AtBeginDocument{%
  \providecommand\BibTeX{{%
    \normalfont B\kern-0.5em{\scshape i\kern-0.25em b}\kern-0.8em\TeX}}}

\setcopyright{acmcopyright}
\copyrightyear{2020}
\acmYear{2020}
\acmDOI{doi}

\acmConference[CONF '20]{Conference}{March 2020}{Chicago, IL}
\acmBooktitle{Conference, March 2020, Chicago, IL}
\acmPrice{15.00}
\acmISBN{isbn}

\usepackage{tabularx}
\begin{document}

\title{To Tweet or Not to Tweet: Covertly Manipulating a Twitter Debate on Vaccines Using Malware-Induced Misperceptions}

\author{Filipo Sharevski}
\affiliation{%
  \institution{DePaul University}
  \streetaddress{243 S Wabash Ave}
  \city{Chicago}
  \state{IL}
  \postcode{60604}
}
\email{fsharevs@cdm.depaul.edu}

\author{Peter Jachim}
\affiliation{%
  \institution{DePaul University}
  \streetaddress{243 S Wabash Ave}
  \city{Chicago}
  \state{IL}
  \postcode{60604}
}
\email{pjachim@depaul.edu}

\author{Kevin Florek}
\affiliation{%
  \institution{DePaul University}
  \streetaddress{243 S Wabash Ave}
  \city{Chicago}
  \state{IL}
  \postcode{60604}
}
\email{kflorek@depaul.edu}

\renewcommand{\shortauthors}{F. Sharevski, P.Jachim, K. Florek}

\begin{abstract}
Trolling and social bots have been proven as powerful tactics for manipulating the public opinion and sowing discord among Twitter users. This effort requires substantial content fabrication and account coordination to evade Twitter's detection of nefarious platform use. In this paper we explore an alternative tactic for covert social media interference by inducing misperceptions about genuine, non-trolling content from verified users. This tactic uses a malware that covertly manipulates targeted words, hashtags, and Twitter metrics before the genuine content is presented to a targeted user in a covert man-in-the-middle fashion. Early tests of the malware found that it is capable of achieving a similar goal as trolls and social bots, that is, silencing or provoking social media users to express their opinion in polarized debates on social media. Following this, we conducted experimental tests in controlled settings ($N=315$) where the malware covertly manipulated the perception in a Twitter debate on the risk of vaccines causing autism. The empirical results demonstrate that inducing misperception is an effective tactic to silence users on Twitter when debating polarizing issues like vaccines. We used the findings to propose a solution for countering the effect of the malware-induced misperception that could also be used against trolls and social bots on Twitter. 
\end{abstract}




\maketitle


\section{Introduction}
Social media, beyond networking people, has become an important source for news dissemination and public discourse \cite{Badawy}, \cite{Kirman}. Platforms like Twitter and Facebook are the go-to places where most of the people monitor the public opinion and participate in debates on issues with a high moral component like politics, minority rights, vaccines, gun control, or immigration \cite{Thompson}, \cite{DiResta}. A diversity of malicious actors, exploiting the openness and lax content/user verification of the platforms, regularly promulgate fabricated and inflammatory content with the goal to manipulate the public opinion and sow discord between the people participating in polarized debates \cite {Spangher}. The commonly used methods are: (1) \textit{trolling} -  where a user “constructs the identity of sincerely wishing to be part of a debate but whose real intentions are to cause disruption;" \cite{Hardaker}; or (2) \textit{social botnets} - where linked bot accounts tweet and retweet messages in unison to create misperception of consensus on a polarized topic or skew metrics of popularity and reach" \cite{Cresci}. 


Malicious actors in the past utilized these methods to create misperceptions in political debates. During the US elections in 2016, state-sponsored trolls infused both pro-Trump (Russian) and anti-Trump (Iranian) fabricated and inflammatory content \cite{Zannettou}. In a similar way, content with an anti-EU sentiment was disseminated during the Brexit campaign \cite{Llewellyn}. The troll accounts and social bots were found to interfere with the \#BlackLivesMatter movement \cite{Stewart} and the gun control debate on Twitter \cite{Danai}. A significant trolling and social bot effort was also invested in a coordinated tweeting and retweeting campaign to amplify the vaccines debate on Twitter \cite{Broniatowski}. Aware of this nefarious use, Facebook and Twitter began to remove trolling content and delete bot accounts \cite{twitter}, \cite{fb}.  

The malicious actors will likely continue to search for covert alternatives to manipulate public opinion and sow discord through social media while remaining undetected. To evade trolling detection and yet cause disruption in a debate, an option is to create content with modified linguistic features that differ from known trolling profiles and posting/tweeting behaviour \cite{Ghanem}, \cite{Addawood}. To evade social bot detection and yet induce misperceptions, an option is to coordinate the posting/tweeting activity to resemble a collective behavior characteristic for humans instead of (semi)automated bots \cite{Luceri}, \cite{Duh}, \cite{Bello}. These and similar options seem logical and might be expected to emerge in the next wave of trolling and social bot activity. However, they all require substantial production and dissemination of content and management of bot accounts. 

In this paper we explore an alternative for covert manipulation of social media debates by inducing misperceptions about genuine, non-trolling content from verified users. Instead of crafting content and coordinating accounts, the misperception is induced by a malware that acts as a man-in-the-middle between the social media platform and a user and manipulates how authentic content is \textit{perceived} by the user themself. Studies on manipulating online information point that \textit{induced misperceptions} represent an effort of a malicious actor to "lead an individual towards making false or implausible interpretations of a set of true facts" \cite{Benkler}. In the same manner, this malware covertly swaps, rearranges, or removes words, hashtags, or metrics presented to an individual to induce interpretation of a set of true facts to the objective of a malicious actor. Using a malware to induce misperception, to our knowledge, is a zero-day exploit because it allows the targeted user to verify the authenticity of a post or tweet, thus bypassing all conventional cues people use to detect trolling or fabricated content \cite{Ferreira}.

We tested the effect of the misperception-inducing malware in controlled settings on a Twitter debate on the risk of vaccines causing autism. The goal was to investigate whether this malware can be used to engineer the \textit{spiral-of-silence} effect on social media, e.g. manipulate how users perceive genuine tweets and accordingly express or silence their opinion. Spiral-of-silence theory argues that if people perceive that their own opinion is in the minority, they are less likely to share it in a debate, especially when discussing issues with high moral component  \cite{Matthes}. A sample of 315 participants was exposed to a polarized debate on the risks of vaccines causing autism. We used the malware to engineer a perception that a larger-than-expected twitter population share the anti-vaccine sentiment \cite{Blankenship}. The malware was packaged as a web browser extension as a low-cost option that allowed controlled use only in laboratory settings (alternative packaging is also discussed in the paper) \cite{Newman}. The results show that the malware could successfully engineer the spiral-of-silence effect for pro-vaccine Twitter users, "nudging" most of them to refrain from sharing their personal opinion or endorsing an account perceived as overtly anti-vaccine. Since there is an effort to aid users to counter social bots and trolls on social media \cite{Cheng}, we further investigated how a suggested, ready-made response could help Twitter users disrupt the spiral-of-silence effect. The results show that users welcome such an aid and prefer a response attacking the \textit{authority} of the anti-vaccine sentiment. 

This paper proceeds as follows: Section 2 provides the theoretical background of the spiral-of-silence effect and its materialization in the social media landscape; Section 3 elaborates on the concept of malware-induced misperception; Section 4 details the use of malware induced-misperception to engineer the spiral-of-silence effect on a Twitter debate regarding the risk of vaccines causing autism in controlled settings; Section 5 reports the empirical results of the study; The implications of the results for the next wave of trolling and social bot activity on social media are discussed in Section 6; Finally, Section 7 concludes the paper.  


\section{Spiral-of-Silence}
\subsection{Theoretical Background}
The spiral-of-silence theory posits that people whose opinions do not coincide with the majority opinion, as they perceive it, tend to silence themselves fearing social isolation \cite{Noelle-Neumann}. These opinions are usually on issues with a high moral component, e.g. politics, public health, minority rights, immigration, or abortion \cite{Scheufle}. The effect of the spiral-of-silence is more likely to be maximized when an issue that is controversial and morally relevant receives a great deal of media coverage \cite{Newman}. Therefore, the theory posits that people use their media environment to alert themselves about the perceived appropriateness of publicly expressing their opinions.

Historically, the spiral-of-silence was mostly investigated on political issues with a limited focus on health or social issues \cite{Hayes}, \cite{McKeever}. When discussing non-political issues, studies have found that the spiral-of-silence effect does not always materialize in the same way as in political debates and discourse \cite{Scheufle}, \cite{LinSalawen}, \cite{Matthes}. That is, people don't use the majority opinion congruence as the only decisive factor for expressing their own opinion. For example, the issue importance was found to be a factor predicting speaking out on the topics of abortion and immigration \cite{Matthes}. Another factor influencing the willingness to express one's opinion is their attitude certainty on the polarized topic. Since our study focuses on vaccines as a health issue with a high moral component, we considered the "issue importance" and "attitude certainty" factors when testing for the possibility of engineering the spiral-of-silence effect with a misperception-inducing malware.

\subsection{Spiral-of-Silence Online} 
The spiral-of-silence theory, developed for face-to-face communication, originally considered printed and televised mass media content. A true consensus on the majority public opinion was easier to build back then and, thus also easier perceive because of limited choices for media consumption or opinion expression. The Internet changed the way people communicate by providing anonymity and selectivity, access to diverse media content, and choices where and with whom to share their opinion \cite{Gearhart2}. This change prompted tests of the spiral-of-silence theory in the context of online communication. Authors in \cite{Kim} suggest that willingness to express one's opinion online is also influenced by the issue importance next to the majority opinion congruence, proving that this factor is also relevant for the spiral-of-silence online. Authors in \cite{Nekmat} have found that the online environment affords people not only to explicitly express an opinion (e.g. by writing comments in online forums, social media, or websites), but also to take actions to implicitly communicate their stance (e.g. reposting, liking, or joining someone's else opinion). Because our study uses Twitter as an online communication platform that affords retweeting, liking, or following a tweet/account, we took into consideration both aspects of online opinion expression into account when analyzing the potential of using a misperception-inducing malware to engineer the spiral-of-silence effect online. 

\subsection{Spiral-of-Silence on Social Media}
Social media interactions are anchored in real-world relationships and people on social media express their opinions in ways to avoid "appearing unpopular or undesirable within the social media community" \cite{Matthes}. Confirming the grounds for the existence of the spiral-of-silence effect on social media, authors in \cite{Chun}, \cite{Chan}, \cite{Hoffmann}, \cite{Hampton} and \cite{Gearhart} demonstrated the robustness between the perceived opinion congruence and the explicit opinion expression on political and non-political topics debated on social media. For example, authors in \cite{Kushin} assessed the spiral-of-silence on Facebook in the context of the 2016 US presidential election. Their analysis suggests that the more people perceived public opinion support for Hillary Clinton, the less likely they were to share a divergent opinion on the platform. Another study investigating the spiral-of-silence on Facebook on the issue of freedom of speech on college campuses has found that the perceived opinion congruence as well as the issue importance, played a decisive role in one's willingness to express their opinion publicly [x].

The spiral-of-silence effect is evidenced on Twitter too, where a study exploring people's agreement with/opposition to nuclear power plants has found that users who recognized that their own opinion is in the majority had a positive effect on the number of tweets they tweeted \cite{Lee}. Despite this study, and perhaps few other ones on issues with a lesser moral component (e.g. \cite{Lee1} or \cite{Wang}), the investigation of the spiral-of-silence effect on social media was predominately focused on Facebook as a platform of choice. To address this gap, we focused on Twitter as a social media platform of choice. We also selected Twitter because it is the go-to place for vaccines debate and as such makes a relevant platform for testing any misperceptions regarding the vaccine debate effectiveness induced by a malware \cite{Broniatowski}. The Twitter interface also has a unique set of affordances for one to both explicitly (e.g. tweeting) and implicitly (e.g. retweet, like, block, follow) express their opinion, providing us the opportunity for a more nuanced test of the effect of the malware in engineering the spiral-of-silence effect on social media.

\section {Malware-Induced Misperception}
\subsection{Concept}
Distorting an individual's map of reality by inducing misperception has become a significant problem on social media \cite{Benkler}. Malicious actors using trolls and social bots flooded Facebook and Twitter with rumors, fabricated content, and inflammatory comments to bias people and sway their votes prior to the US presidential election in 2016 \cite{Badawy}, \cite{Spangher}. They also engaged in strategic infusion of fabricated content for issues with high moral component, such as the risks of e-cigarettes or the link between the vaccines and autism \cite{Jamison}. The idea was to "manipulate the perception of public opinion and sow discord between people debating health issue topics" in order to perpetuate a latent state of disagreement among the American public \cite{Cresci}. For this purpose the malicious actors used a considerable number of trolls and bot accounts who infused fabricated and inflammatory content in a coordinated fashion. 

The concept of \textit{malware-induced misperception} is inspired by these efforts but replaces the need for fabricating content or infusing inflammatory tweets and comments. The malware also elevates worries that the social media platform can detect a misperception campaign. Instead, the misperception takes place on a local machine or smartphone where the malware covertly rearranges words, endorsement actions (e.g. likes, or shares), and topic keywords (e.g. hashtags) of a genuine social media post while the targeted user is reading it in real time. Studies on manipulating public opinion point that \textit{induced misperceptions} are efforts of a malicious actor to "lead an individual towards making false or implausible interpretations of a set of true facts" \cite{Benkler}. By targeting genuine content, the malware allows the targeted individual to verify the facts and the credibility of a source, thus bypassing all conventional cues people use to detect "phishy" content \cite{Ferreira}. The goal of the malware is to covertly manipulate the data in transit and induce interpretation of genuine content  towards the objective of the malicious actor. 

\subsection{Implementation}
The misperception-inducing malware can be packaged as a browser extension (e.g. Chrome) or a third party custom Twitter application. The malware usually is disguised as seemingly benign to lure a user to install it in the first place. The social engineering persuasion through disguise is needed because the malware requires text manipulation permissions that later will be leveraged for implementation of the misperception-inducing logic \cite{Vincent}. Developing third-party extensions and apps is free and a benign software can pass all the security checks before publishing \cite{Newman}. For example, a browser extension variant of the malware can disguise the misperception-inducing logic and pass the security checks by posing as an "accessibility (a11y) extension" that claims the rewording is done to help non-native English speakers understand English slang on social media \cite{Jang}. The malware could be packaged as a third-party smartphone app that, for example, provides user-tailored Twitter experience by filtering hashtags, content, and users ~\cite{Seals}.

The malware implements a word/number replacement logic if a target word/number is detected on the Twitter page. The malware parses the Twitter content with a \texttt{findMatch()} function to detect a potential and returns the opposite valenced word/number. A \texttt{textSwap()} function then replaces the occurrences of the initially detected word/number based on a configurable logic (all occurrences, only the first occurrence, or only if the occurrence is in the comments section of a Twitter page). This is the simplest, low cost and low complexity version of the malware. A malicious actor can implement more complex logic where the linguistic manipulation can take place only in certain parts of the Twitter content or only in Twitter posts from a specific person or on a particular issue, for example, only posts with the hashtag "\#vaccines" but not other hashtags. The string array of "valence words and numbers" need not to be predefined in that a malicious actor could use natural language processing and activity analytics to analyze authentic Twitter content and adapt the rearrangement that makes the most sense in the context of target individuals' Twitter diet [x]. Using a Markov chain model can be trained to choose replacement words or numbers based on an identified corpus of Twitter content \cite{Downey}. 

\subsection{Pilot Test}
For the purpose of our study we developed the malware as a browser extension in JavaScript as a more economic proof-of-concept variant. We conducted a pilot study with 24 volunteer participants where we tested the malware's potential to induce misperception on a polarized tweet. All participants were 18 years or older, interacted on Twitter through a web browser, and had prior knowledge of past social media trolling, misperception, and fake news campaigns. The preliminary question was to gauge whether participants are open to using browser extensions for improved browsing experience \cite{Vincent}. Most of them responded they already do use various extensions that improve their productivity. Some were aware that browser extensions could contain spyware and steal personal information, so they look for legitimate extensions only on the browser application stores. Few were aware of extensions that manipulate content, like the Twitter demetricator that hides the number of likes, retweets, and replies to enable a more immersive interaction \cite{Grosser}. None of them were aware of browser extensions that covertly rearrange the Twitter content before it is rendered in a browser. This was important feedback suggesting that it is plausible for a malicious actor to employ a legitimacy-by-design to persuade the target user to install a browser extension in the first place \cite{Newman}. 

The pilot participants first encountered a genuine Twitter post shown in Figure 1 that we adopted from \cite{Mitra}. We asked their "attitude certainty" and how important the issue of vaccines alleged link to autism is. Based on that,  we then asked what action they would take if they see this post on their Twitter feed. Most of them reported that they are certainly pro-vaccine,that the issue is very important, and that they would probably retweet or possibly provide a short comment saying "Yeah, vaccines work" or "Vaccines created adults; without them we wouldn't grow up to a decent age." Those that expressed anti-vaccine sentiment weren't as certain, indicating they would probably ignore the tweet. 

\begin{figure}[htb]
\centering
  \includegraphics[width=0.9\columnwidth]{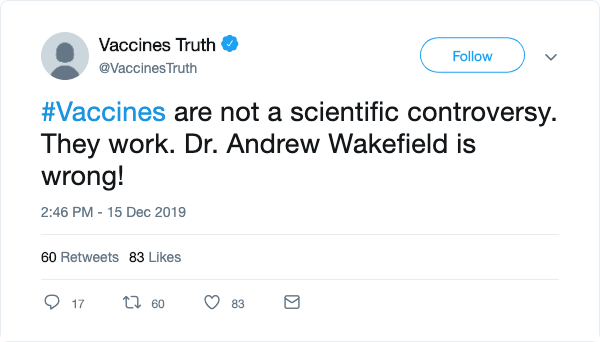}
  \caption{malware extension "off"}
  \label{Fig2}
\end{figure}

Then we used the malware to covertly remove the word "not," insert the word "don't" and swap the word "wrong" with "right" as shown in Figure 2. We also covertly doubled the number of retweets, replies, and likes to induce misperception that there is growing consensus on the anti-vaccine side \cite{Jamison}. We asked the pro-vaccine participants what action would they take if they see this new post into their twitter feed. Most of them reported they were not inclined to reply, retweet, nor like it. Some of them indicated they would probably block the account because "it seems like \#fakenews" or "a bunch of trolls." The anti-vaccine participants mostly said they would like the post and possibly retweet it. 

\begin{figure}[htb]
\centering
  \includegraphics[width=0.9\columnwidth]{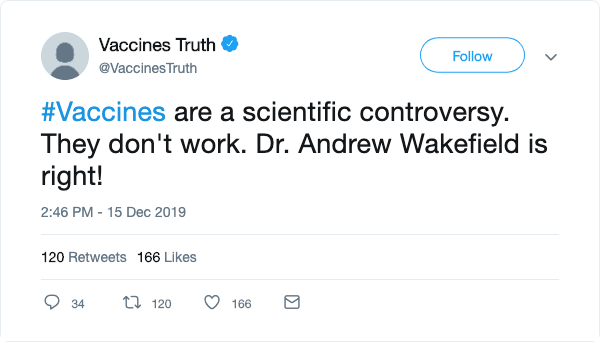}
  \caption{malware extension "on"}
  \label{Fig3}
\end{figure}

An important feature of the malware is that it allowed the participants, aware of trolling and fabricated/inflammatory content, to verify the account (i.e. see the verification icon next to the account name). The outcome of the pilot study suggested that there are plausible grounds to suspect that a malware can induce misperception about a polarized topic, for example a debate on vaccines on Twitter, and with that socially engineer the spiral-of-silence effect. This motivated us to explore the effects of this micro-targeting alternative for public opinion manipulation with a larger sample.


\section {Engineering a Spiral-of-Silence Effect on Twitter}
Social media platforms have been found to actively shape users’ perceptions of minority and majority opinion climates with their content curation algorithms, and with that, "engineer" the spiral-of-silence at scale by manipulating what outside content is presented to individual users \cite {Morgan}, \cite{Grover}. In the past they allowed for fabricated and inflammatory content to be infused by state-controlled trolls and social bots that aimed to distort users’ perceptions of minority and majority opinion climates about polarizing issues \cite{Kushin}.  Recently, an emerging line of research is looking into how the spiral-of-silence effect can be "engineered" on social media as an alternative to the platforms' news feed algorithms or state-controlled trolls and bots [x], [x]. Instead of \textit{curating} or infusing \textit{fabricated/inflammatory} content, the idea is to induce misperception about \textit{genuine} content using a malware that covertly alters targeted words/numbers before the post is rendered to the targeted user. In this study, we utilized the same approach to explore how a similar malware can be used to induce misperception about genuine Twitter content, and with that, affect one's willingness to express their opinion. As a polarizing issue, we chose a Twitter debate on vaccines and the risk of causing autism, following the reports about Russian trolling activity aimed to amplify this debate and sow discord \cite{Broniatowski}.


\subsection{Research Questions}
To explore the possibility of socially engineering a spiral-of-silence effect in a Twitter debate on vaccines' effectiveness, we set to answer the following research questions: \\

\textbf{Research Question 1a}: \textit{How would a malware-induced misperception about the effectiveness of vaccines affect one's choice of 'opinion expression strategies' on Twitter, either a personally crafted tweet, a suggested tweet, or no tweet at all?}\\

\textbf{Research Question 1b}: \textit{How would a suggested opinion expression strategy, e.g. a recommended set of readily available tweets, help one disrupt the spiral-of-silence effect on Twitter when participating in a debate on vaccines' effectiveness?}\\

\textbf{Research Question 2}: \textit{How would a malware-induced misperception about the effectiveness of vaccines affect one's choice of 'opinion expression actions' on Twitter, either retweet, like, block, start following, or ignore a given tweet/account?}\\

\textbf{Research Question 3}: \textit{How is one's choice of opinion expression strategies and actions on Twitter influenced by their authentic representation when responding to a malware-manipulated tweet about the effectiveness of vaccines?}

\subsection{Study Design}
The study utilized an original pro-vaccine tweet shown in Figure 1 and a malware manipulated tweet shown in Figure 2. The original tweet was based on the tropes commonly used to argue vaccination effectiveness elaborated in \cite{Kata}. The tweet had two hashtags that are popular in the vaccines debate and came from a generic "@Vaccines-Truth" account \cite{Broniatowski}. We used such an approach to capture the initial reaction to a "new" tweet that was based on real, authentic anti-vaccine rhetoric \cite{Orgaz}. We used the malware to manipulate: (1) the content of the tweet; (2) the wording of the hashtags; and (3) the metrics of the tweet. The malware covertly replaced the word "Many" with the word "No" from the original, \textit{pro-vaccine} tweet shown in Figure 1 to make it appear as the \textit{anti-vaccine} tweet shown in Figure 2. The original pro-vaccine hashtags \#provax and \#vaccineswork were covertly replaced with \#antivax and \#vaccinesdontwork. The malware also quadrupled the total number of comment tweets, number of retweets, and number or likes to manipulate the perception of the majority opinion and with that to covertly  "engineer" spiral-of-silence conditions. The malware therefore changed the number the comments to $32$,the number of retweets to $160$ and the number of likes to $548$ (the numbers come from an internal analysis of the average metrics and information sharing patterns on Twitter for vaccines-related hashtags \cite{Love}, \cite{Orgaz}). 

\begin{figure}[!h]
\centering
  \includegraphics[width=0.9\columnwidth]{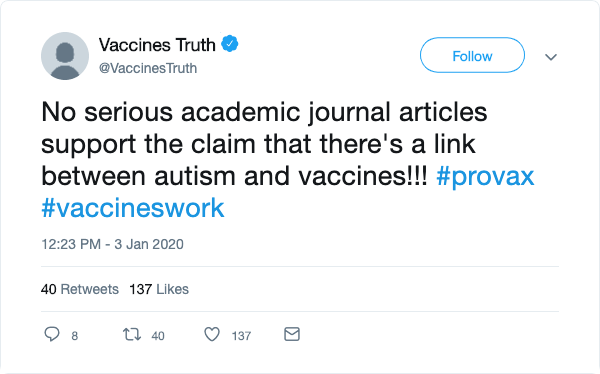}
  \caption{The original tweet with pro-vaccine sentiment.}
  \label{Fig2}
\end{figure}

\begin{figure}[!h]
\centering
  \includegraphics[width=0.9\columnwidth]{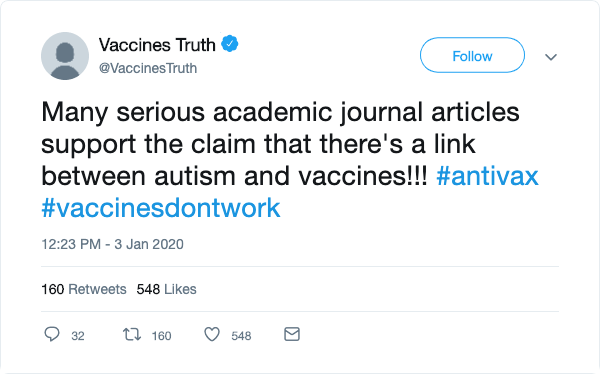}
  \caption{Malware-manipulated tweet to induce anti-vaccine sentiment misperception.}
  \label{Fig2}
\end{figure}

Due to IRB restrictions, we conducted the study using a survey that presented the tweets in a controlled web browser and solicited questions from our participants. This was done to eliminate the need for the participants to respond from their accounts given that the study was anonymous, and didn't collected any personally identifiable information. This set up also enabled us to toggle the malware extension without the risk that the participants could get a hold of the misperception-inducing logic and code. The objective of our study is not to disseminate or encourage a development of similar browser extensions but to understand their effect, given that they provide a very low cost approach for man-in-the-middle manipulation of online information (we discuss the ethics of our study later in the paper). The participants were grouped based on their "attitude certainty" and "issue importance" as either "pro-vaccine, "anti-vaccine," or "ambivalent." The pro-vaccine and ambivalent participants were presented the tweet in Figure 3 and the anti-vaccine tweet in Figure 4. We decided to show an anti-vaccine post to the participants who identified as "ambivalent" to eliminate a possibility that they will simply conform with the dominant pro-vaccine sentiment on Twitter \cite{Blankenship}. 

We modified the notion of opinion expression strategies extensively used in the spiral-of-silence research to fit the Twitter interface and user interaction, i.e. the act of "tweeting" \cite{Hayes}, \cite{Kim}. We also considered the fact that people who debate the vaccines' effectiveness on Twitter regularly borrow lines or share links from other sources \cite{Addawood}. Accordingly, we utilized the following opinion expression strategies to answer the first research question: (1) self-response, where people were left to express their own personal opinion or \textit{tweet themselves}; (2) a suggested response, where people were offered to choose three response tweets from the tropes of vaccine debate provided in \cite{Kata}; and (3) no response at all. The suggested tweets were crafted to appeal to the three most dominant principles of persuasive rhetoric: \textit{authority}, \textit{social proof}, and \textit{labeling} \cite{Cialdini}, \cite{Liu}. For the self-responding participants, we utilized the Linguistic Inquiry and Word Count (LIWC) tool \cite{Pennebaker} for computational modeling of the response language to further understand what kind of response the misperception-inducing malware could provoke. A similar approach for the modeling of vaccines' stigmatized beliefs on social media has been utilized in \cite{Straton} and \cite{Faasse}. Because a Twitter user can take additional endorsement actions as a result of self-expression on Twitter \cite{Kiran}, we utilized the following \textit{opinion expression actions} one can take after reading a tweet to answer the second research question: (1) retweet it; (2) like it; (3) block the account that tweeted it; (3) start following the account that twitted it; (3) or ignore the account/tweet completely. The likelihood was measured on a 7-point Likert scale from "1-extremely likely" to "7-extremely unlikely." To answer the third research question, a set of demographic questions concluded the survey by asking the participants to provide their age, gender, and level of education.


\section{Study Results}
Following an IRB approval for the study, we fielded an online survey ($N =315$) in the period of January 30th to March 4st. The sample was skewed towards the pro-vaccine: 260 or 82.5\% participants indicated they were pro-vaccine, 32 or 10.2\% were anti-vaccine, and 23 or 7.3\% were ambivalent. Six age brackets in the sample were distributed as: 23.5\% [18-22], 22.2\% [23-27], 17.8\% [28-32], 8.3\% [33-37], 4.7\% [38-42], and 23.5\% [43 or more]. The sample was representative with 175 or 55.6\% identified as female, 115 or 36.5\% as male, and 25 or 7.9\% as non-cis individuals (transgender male, transgender female, gender variant/non-conforming, not listed, or preferred not to answer).  The level of education was distributed as: 32 or 10.2\% had a high school degree or equivalent, 74 or 23.5\% had some college but no degree, 128 or 40.6\% had a college degree, and 81 or 25.7\% had a graduate degree.

\subsection{Research Question 1}
The first research question preliminary explored the effect of a covertly manipulated tweet with a misperception-inducing malware on one's choice of opinion expression strategies. Table 1 shows that when the pro-vaccine participants were exposed to an anti-vaccine climate of opinion, only 28.8\% opted out to tweet themselves. When the anti-vaccine group of participants was exposed to a pro-vaccine climate of opinion, only 15.6\% opted out to express their personal opinion. Only one of the ambivalent participants choose to craft their personal tweet. Overall, these results confirm the previous findings that the spiral-of-silence can be socially engineered using a covert manipulation of the content of a controversial social media post [x]. Misperceiving a majority of opposing opinion climate is sufficient to nudge someone to remain silent, that is, refrain from tweeting their personal opinion in response. 

\begin{table}[h]
\renewcommand{\arraystretch}{1.0}
\caption{Choice of an Opinion Expression Strategy} 
\label{table_6}
\centering
\begin{tabularx}{\linewidth}{|l|X|X|X|X|X|}
\hline
 & \textbf{Self} & \textbf{Auth} & \textbf{Soc.Pr.} & \textbf{Label} & \textbf{No} \\
\hline
 \textit{Pro} & 75 & 88 & 73 & 21 & 3\\
\hline
 \textit{Anti} & 5 & 8 & 8 & 9 & 2\\
\hline
\textit{Ambv.} & 1 & 4 & 13 & 4 & 1\\
\hline
\textbf{Total} & 81 & 100 & 94 & 34 & 6\\
\hline
\end{tabularx}
\end{table}

The content analysis of the personal opinion tweets using the LIWC tool \cite{Pennebaker}, shown in Table 2, revealed that the pro-vaccine participants show a less "analytical thinking" pattern (score = 39.41) than the anti-vaccine participants (score = 51.42). Both scores are relatively low, but this is to be expected given the limited space for a tweet and hence an expression of a more elaborated opinion. The pro-vaccine participants are much more confident in their position with much higher "clout" compared to the anti-vaccine participants. This result reflects the dominant, pro-vaccine sentiment on Twitter \cite{Blankenship}. The findings also add to the evidence that the people who are not afraid to disrupt the spiral-of-silence and speak out are the ones with "hard core" attitudes towards the controversial issue debated \cite{Newman}, \cite{Gearhart2}. The anti-vaccine participants were way more "authentic" (score = 85.21) and used a more positive tone (score = 68.66) compared to the pro-vaccine participants. We suspect this is the case because the anti-vaccine position is rather on defense and bears the burden to maintain the position in the face of new and compounding evidence of the positive effects of vaccines \cite{Orenstein}. The only ambivalent self-response expressed a rather sceptic personal opinion, tweeting in reply "There are academic journals that support the link; but it depends who's funding the study; it can be looked at as one-sided and biased." 

\begin{table}[h]
\renewcommand{\arraystretch}{1.0}
\renewcommand{\tabcolsep}{2mm}
\caption{Opinion Content Analysis}
\label{table2}
\centering
\begin{tabularx}{\linewidth}{|l|X|X|X|X|}
\hline
 & \textbf{Analytic} & \textbf{Clout} & \textbf{Authentic} & \textbf{Tone} \\
\hline
\textit{Pro} & 39.41 & 60.68 & 19.7 & 45.21 \\
\hline
\textit{Anti} & 51.42 & 1.2 & 85.21 & 68.66 \\
\hline
\end{tabularx}
\end{table}

The second part of the first research question aimed to understand how a suggested opinion expression strategy, e.g. a recommended set of readily available tweets, helps one disrupt the spiral-of-silence effect on Twitter when participating in a debate on vaccines' effectiveness. As shown in Table 1, the recommended set of readily available tweets were a much more attractive strategy for the majority of the pro-vaccine participants. 33.8\% of the participants chose the \textit{authority} option "All experts deny any vaccines-autism link." 28.1\% of the participants chose the \textit{social proof} option "Vaccines saved us!" Only 8.1\% chose the \textit{labeling} option "You are a conspiracy theorist!" Only 1.2\% of the pro-vaccine participants opted to remain completely silent. Even more, this was the case for the anti-vaccine participants with an equal distribution between the \textit{authority} ("Few experts deny any vaccines-autism link"), \textit{social proof} ("Vaccines didn't save us!"), and \textit{labeling} ("You are in a pocket of Big Pharma") responses. More than half of the ambivalent participants chose the anti-vaccine \textit{social proof} to reflect their not so "hard core" attitude and position. All of these are interesting results showing that a simple aid in form of a \textit{chatbot} could be provided to people interested in participating in a Twitter debate to help them share at least some kind of opinion. The design and utility of such a chatbot, based on a popular approach used by Facebook \cite{Frenkel}, is elaborated in the next section.



\subsection{Research Question 2}
The second research question explored the effect of a covertly manipulated tweet with the misperception-inducing malware on one's opinion expression options. The participants were given a choice of five possible endorsement actions and asked to provide the likelihood of taking either one of them on a 7-point Likert scale (1 - "extremely likely", 7-"extremely unlikely"). Comparing between each of the groups, the results of a Kruskal-Wallis test given in Table 3 indicated a statistically significant difference for the first option (retweet the post) and for the last option (follow the account that tweeted the post). The pro-vaccine participants were much less likely to retweet an opposing tweet to their position, and they were much less likely to start following the posting account compared to the other two groups. Corresponding to the previous findings, it follows that the malware is also able to "nudge" the pro-vaccine advocates to refrain from any endorsement action by manipulating the perception of the majority anti-vaccine climate. 


\begin{table}[h]
\renewcommand{\arraystretch}{1.0}
\renewcommand{\tabcolsep}{2mm}
\caption{Opinion Expression Actions - Test Comparison}
\label{table2}
\centering
\begin{tabularx}{\linewidth}{|l|X|X|}
\hline
 & \textbf{Retweet} & \textbf{Follow}  \\
\hline
\textit{$\chi^2$} & 13.622 & 14.573  \\
\hline
\textit{Asymp. Sig.} & .001* & .001* \\
\hline
\multicolumn{3}{l}{$^*p<.05$} \\
\end{tabularx}
\end{table}


\begin{table}[h]
\renewcommand{\arraystretch}{1.0}
\renewcommand{\tabcolsep}{2mm}
\caption{Opinion Expression Actions - Mean and Std. Dev.}
\label{table2}
\centering
\begin{tabularx}{\linewidth}{|l|X|X|}
\hline
\multicolumn{3}{|c|}{pro-vaccine} \\
\hline
 &  \textbf{Retweet} & \textbf{Follow}  \\
\hline
\textit{Mean}  & 5.03; & 5.72  \\
\hline
\textit{Std. Dev.} & 1.98; & 1.97 \\
\hline
\multicolumn{3}{|c|}{anti-vaccine} \\
\hline
\textit{Mean} & 3.77 & 4.58  \\
\hline
\textit{Std. Dev.}  & 2.08 & 1.86 \\
\hline
\multicolumn{3}{|c|}{Ambivalent} \\
\hline
\textit{Mean} & 5.13 & 5.45 \\
\hline
\textit{Std. Dev.} & 1.57 & 1.71 \\
\hline
\end{tabularx}
\end{table}

\subsection{Research Question 3}
\subsubsection{Opinion Expression Strategies}
The third research question explored how one's authentic representation influences the choice of opinion expression strategies and actions when responding to a tweet about the effectiveness of vaccines, manipulated by a misperception-inducing malware. Table 5 provides a breakdown of the choices of expression strategies per age bracket in our sample. The percentage of self-responding pro-vaccine participants is more than 20\% for each age bracket. This breakdown, in line with the findings in the first research question, further suggests that an attacker interested in targeting a twitter population with a misperception-inducing malware stands a roughly equal chance to "silence" every age bracket of users. 

Looking into the suggested opinion expression strategies, the \textit{authority} response was the most popular in all age brackets except for age bracket [42 and older] - they mostly preferred the \textit{social proof} response. This gives more insight into how a malicious actor crafts the misperception logic. The malware targeting 42 and older users could aim to induce misperception that there is a lack of consensus on the vaccines efficiency as we initially did in our study, but in targeting the other age bracket, the logic could be driven towards undermining any vaccine authority beyond academic journals. A similar conclusion holds for the anti-vaccine participants in the [42 and older] bracket, given that they were the most inclined to disrupt the spiral-of-silence with a self-response tweet compared to the other age brackets. 

\begin{table}[h]
\renewcommand{\arraystretch}{1.0}
\caption{Opinion Expression Strategy vs. Age} 
\label{table_6}
\centering
\begin{tabularx}{\linewidth}{|l|X|X|X|X|X|}
\hline
\multicolumn{6}{|c|}{[18-22]} \\
\hline
 & \textbf{Self} & \textbf{Auth} & \textbf{Soc.Pr.} & \textbf{Label} & \textbf{No} \\
\hline
 \textit{Pro} & 19 & 25 & 16 & 6 & 0\\
\hline
 \textit{Anti} & 1 & 4 & 1 & 0 & 1\\
\hline
\textit{Ambivalent} & 0 & 1 & 0 & 0 & 1\\
\hline
\multicolumn{6}{|c|}{[23-27]} \\
\hline
 & \textbf{Self} & \textbf{Auth} & \textbf{Soc.Pr.} & \textbf{Label} & \textbf{No} \\
\hline
 \textit{Pro} & 21 & 21 & 14 & 3 & 0\\
\hline
 \textit{Anti} & 1 & 3 & 2 & 0 & 0\\
\hline
\textit{Ambivalent} & 0 & 0 & 4 & 1 & 0\\
\hline
\multicolumn{6}{|c|}{[28-32]} \\
\hline
 & \textbf{Self} & \textbf{Auth} & \textbf{Soc.Pr.} & \textbf{Label} & \textbf{No} \\
\hline
 \textit{Pro} & 13 & 17 & 12 & 8 & 1\\
\hline
 \textit{Anti} & 0 & 0 & 0 & 1 & 0\\
\hline
\textit{Ambivalent} & 0 & 2 & 2 & 1 & 0\\
\hline
\multicolumn{6}{|c|}{[33-37]} \\
\hline
 & \textbf{Self} & \textbf{Auth} & \textbf{Soc.Pr.} & \textbf{Label} & \textbf{No} \\
\hline
 \textit{Pro} & 7 & 10 & 3 & 3 & 0\\
\hline
 \textit{Anti} & 0 & 0 & 1 & 1 & 0\\
\hline
\textit{Ambivalent} & 1 & 0 & 0 & 0 & 0\\
\hline
\multicolumn{6}{|c|}{[38-42]} \\
\hline
 & \textbf{Self} & \textbf{Auth} & \textbf{Soc.Pr.} & \textbf{Label} & \textbf{No} \\
\hline
 \textit{Pro} & 3 & 4 & 4 & 0 & 4\\
\hline
 \textit{Anti} & 0 & 1 & 0 & 0 & 0\\
\hline
\textit{Ambivalent} & 0 & 0 & 0 & 1 & 0\\
\hline
\multicolumn{6}{|c|}{[42 and older]} \\
\hline
 & \textbf{Self} & \textbf{Auth} & \textbf{Soc.Pr.} & \textbf{Label} & \textbf{No} \\
\hline
 \textit{Pro} & 12 & 11 & 24 & 1 & 1\\
\hline
 \textit{Anti} & 3 & 4 & 4 & 3 & 3\\
\hline
\textit{Ambivalent} & 0 & 1 & 7 & 1 & 0\\

\hline
\end{tabularx}
\end{table}

The content analysis per age bracket given in Table 6 revealed that the least analytical in their responses are the pro-vaccine participants in the age bracket of [33-37],but they are the most confident ones in their responses (score = 79.76). The most analytical are the pro-vaccine participants in the [23-27] (score = 47.43) and [28-32] (score = 47.92) brackets, although with an opposite levels of confidence. The least confident pro-vaccine participants were in the [38-42] (score = 4.8) and [42 and older] (score = 39.15) brackets. The [42 and older] however were the most authentic ones in their responses compared to the other pro-vaccine groups (score = 62.65). The [38-42] showed the most positive tone (score = 92.40) while the [33-37] showed the most negative tone in their responses (score = 8.95). Overall, the analytical thinking is relatively low, but as we noted above, that is probably due to the limited word count for writing a tweet. The clout is relatively high for the younger participants. Except the [42 and older], the other age brackets weren't particularly authentic in their responses. The tone varies per category with no particular pattern. This result suggests that the misperception-inducing malware is capable of provoking a stark, to-the-point, and often negative tone self-response as a way of disrupting the spiral-of-silence in a vaccine debate on Twitter. The content analysis for the anti-vaccine and ambivalent participants per age was the same as the one in section 5.1 (not included in the table).  

\begin{table}[h]
\renewcommand{\arraystretch}{1.0}
\renewcommand{\tabcolsep}{2mm}
\caption{Opinion Content Analysis per Age Bracket}
\label{table2}
\centering
\begin{tabularx}{\linewidth}{|l|X|X|X|X|}
\hline
\multicolumn{5}{|c|}{pro-vaccine} \\
\hline
 & \textbf{Analytic} & \textbf{Clout} & \textbf{Authentic} & \textbf{Tone} \\
\hline
\textit{18-22} & 37.20 & 76.16 &	26.07 &	40.86 \\
\hline
\textit{23-27} & 47.43&	72.17&	16.38&	33.00 \\
\hline
\textit{28-32} & 47.92 &	47.20&	8.05&	77.29 \\
\hline
\textit{33-37} & 24.62	& 79.76	& 3.01	& 8.95 \\
\hline
\textit{38-42} & 27.87	&4.80&	14.24&	92.40 \\
\hline
\textit{42+} & 31.95&	39.15&	62.65&	42.41 \\
\hline
\end{tabularx}
\end{table}

Table 7 provides a breakdown of opinion expression strategies per gender identity in our sample. The pro-vaccine male participants were more inclined to self-express than the female and non-cis participants. The pro-vaccine male participants also preferred the \textit{authority} response while the female and non-cis participants were equally interested in both the \textit{authority} and \textit{social proof} responses. It follows that the malware can particularly target female and non-cis users to induce misperception with various logic and nudge them to silence their opinion.  

Interestingly, the anti-vaccine and ambivalent female participants were more inclined to self-respond to the pro-vaccine tweet compared to the male and non-cis participants. From the malware-inducing misperception perspective, the results suggest that anti-vaccine females could be nudged to disrupt the spiral-of-silence as a result of their "hard core" attitudes and beliefs in their position if the the malware was used to make an originally anti-vaccine tweet appear as it is pro-vaccine (opposite of our study). Such a predisposition has been a predictor for similar behaviour in studies concerning the spiral-of-silence effect on social media [x], \cite{Gearhart2}. 

\begin{table}[h]
\renewcommand{\arraystretch}{1.0}
\caption{Opinion Expression Strategy vs. Gender} 
\label{table_6}
\centering
\begin{tabularx}{\linewidth}{|l|X|X|X|X|X|}
\hline
\multicolumn{6}{|c|}{Female} \\
\hline
 & \textbf{Self} & \textbf{Auth} & \textbf{Soc.Pr.} & \textbf{Label} & \textbf{No} \\
\hline
 \textit{Pro} & 34 & 49 & 46 & 10 & 2\\
\hline
 \textit{Anti} & 4 & 3 & 3 & 4 & 2\\
\hline
\textit{Ambivalent} & 1 & 3 & 10 & 3 & 0\\
\hline
\multicolumn{6}{|c|}{Male} \\
\hline
 & \textbf{Self} & \textbf{Auth} & \textbf{Soc.Pr.} & \textbf{Label} & \textbf{No} \\
\hline
 \textit{Pro} & 36 & 34 & 23 & 9 & 0\\
\hline
 \textit{Anti} & 1 & 2 & 3 & 4 & 0\\
\hline
\textit{Ambivalent} & 0 & 1 & 2 & 0 & 0\\
\hline
\multicolumn{6}{|c|}{Non-Cis} \\
\hline
 & \textbf{Self} & \textbf{Auth} & \textbf{Soc.Pr.} & \textbf{Label} & \textbf{No} \\
\hline
 \textit{Pro} & 3 & 4 & 4 & 2 & 1\\
\hline
 \textit{Anti} & 0 & 2 & 1 & 1 & 0\\
\hline
\textit{Ambivalent} & 0 & 0 & 1 & 1 & 0\\
\hline
\end{tabularx}
\end{table}

The content analysis per gender identity in Table 8 revealed that the least analytical are the pro-vaccine female participants (score =30.86) while the most analytical are the non-cis participants (score = 44.96). The non-cis pro-vaccine participants are the most confident in sharing their opinion (score = 91.35), but they are less authentic than male pro-vaccine participants (score = 20.26) and less positive in tone (score = 48.81). These findings confirm the provoking effect of the malware on the willingness to express one's opinion on social media when controlling for gender identity [x]. 

We have included the anti-vaccine content analysis only for the female participants (the anti-vaccine male/non-cis and the ambivalent participants responses were missing or were the same as in section 5.1). The anti-vaccine female participants were more analytical than any pro-vaccine participants (score = 50.06), much less confident (score = 1.00), far more authentic (score = 87.1) and responded in a much more positive tone (score = 95.29). This result also gives further insight into our previous observation of "hard core" anti-vaccine beliefs. It appears that to defend this position, one needs to be more analytical, more positive, and more authentic if they are to compensate for the eroded confidence in face of new evidence about vaccines' effectiveness \cite {Tomeny}.  

\begin{table}[h]
\renewcommand{\arraystretch}{1.0}
\renewcommand{\tabcolsep}{2mm}
\caption{Opinion Content Analysis per Gender Identity}
\label{table2}
\centering
\begin{tabularx}{\linewidth}{|l|X|X|X|X|}
\hline
\multicolumn{5}{|c|}{pro-vaccine} \\
\hline
 & \textbf{Analytic} & \textbf{Clout} & \textbf{Authentic} & \textbf{Tone} \\
\hline
\textit{Female} & 30.68&	65.97&	9.44&	34.71 \\
\hline
\textit{Male} & 39.12 &	58.34	&20.26&	48.81 \\
\hline
\textit{Non-Cis} & 44.96&	91.35&	11.00&	3.73 \\
\hline
\multicolumn{5}{|c|}{anti-vaccine} \\
\hline
 & \textbf{Analytic} & \textbf{Clout} & \textbf{Authentic} & \textbf{Tone} \\
\hline
\textit{Female} & 50.06& 1.00 &	87.51	& 95.29 \\
\hline
\end{tabularx}
\end{table}

Table 9 provides a breakdown of the opinion expressions strategies per level of education in our sample. The participants with higher degrees were more inclined to self-respond, regardless of their position on the vaccines' effectiveness. The pro-vaccine participants with a high school degree and graduate degree preferred the \textit{social proof} response over the \textit{authority} response compared to the other groups. From a misperception-inducing perspective, these results suggest that level of education influences the targeting users with the malware. This susceptibility to a vaccine sentiment based on one's level of education was also observed in ~\cite{Tomeny}. 

\begin{table}[h]
\renewcommand{\arraystretch}{1.0}
\caption{Opinion Expression Strategy vs. Education} 
\label{table_6}
\centering
\begin{tabularx}{\linewidth}{|l|X|X|X|X|X|}
\hline
\multicolumn{6}{|c|}{High school degree or equivalent} \\
\hline
 & \textbf{Self} & \textbf{Auth} & \textbf{Soc.Pr.} & \textbf{Label} & \textbf{No} \\
\hline
 \textit{Pro} & 4 & 6 & 11 & 0 & 1\\
\hline
 \textit{Anti} & 0 & 4 & 1 & 1 & 0\\
\hline
\textit{Ambivalent} & 0 & 1 & 3 & 0 & 0\\
\hline
\multicolumn{6}{|c|}{Some college but no degree} \\
\hline
 & \textbf{Self} & \textbf{Auth} & \textbf{Soc.Pr.} & \textbf{Label} & \textbf{No} \\
\hline
 \textit{Pro} & 18 & 23 & 11 &9 & 1\\
\hline
 \textit{Anti} & 0 & 0 & 3 & 2 & 1\\
\hline
\textit{Ambivalent} & 0 & 1 & 4 & 0 & 1\\
\hline
\multicolumn{6}{|c|}{College degree} \\
\hline
 & \textbf{Self} & \textbf{Auth} & \textbf{Soc.Pr.} & \textbf{Label} & \textbf{No} \\
\hline
 \textit{Pro} & 38 & 42 & 25 & 6 & 1\\
\hline
 \textit{Anti} & 2 & 3 & 1 & 3 & 1\\
\hline
\textit{Ambivalent} & 0 & 1 & 3 & 1 & 0\\
\hline
\multicolumn{6}{|c|}{Graduate degree} \\
\hline
 & \textbf{Self} & \textbf{Auth} & \textbf{Soc.Pr.} & \textbf{Label} & \textbf{No} \\
\hline
 \textit{Pro} & 15 & 17 & 25 & 6 & 0\\
\hline
 \textit{Anti} & 2 & 1 & 3 & 3 & 0\\
\hline
\textit{Ambivalent} & 1 & 1 & 3 & 3 & 0\\
\hline
\end{tabularx}
\end{table}

The content analysis per level of education in Table 10 reveals that the least analytical and least confident were the pro-vaccine participants with a graduate degree. They maintained a neutral tone and a rather low level of authenticity. We suspect that this might be due to the their "unworthy debate" stance, which has been observed to be the case in polarized debates on Twitter \cite{Yang}. However, this result shows that the malware could potentially nudge the highly educated, on-the-fence users to resort to silence by implementing a logic that shows a rather rigid and dismissive anti-vaccine sentiment \cite{Love}. In contrast, the most analytical are the anti-vaccine participants with college and graduate degrees. Although the tweeting confidence for these groups of participants is very low (scores = 23.75, 1.07, and 2.31, respectively), as we have seen before, they tend to be the most authentic (score = 88.50) and use a positive tone (scores = 97.19 and 73.64, respectively).

\begin{table}[h]
\renewcommand{\arraystretch}{1.0}
\renewcommand{\tabcolsep}{2mm}
\caption{Opinion Content Analysis per Level of Education}
\label{table2}
\centering
\begin{tabularx}{\linewidth}{|l|X|X|X|X|}
\hline
\multicolumn{5}{|c|}{pro-vaccine} \\
\hline
 & \textbf{Analytic} & \textbf{Clout} & \textbf{Authentic} & \textbf{Tone} \\
\hline
\textit{HS} & 40.38	 & 39.87 &	63.54 &	73.64 \\
\hline
\textit{SC} & 32.01 & 	77.33 &	21.42 &	25.77 \\
\hline
\textit{C} & 46.64 &	60.28 &	12.51 &	50.03 \\
\hline
\textit{G} & 19.76 & 	23.75 &	34.27 &	52.57 \\
\hline
\multicolumn{5}{|c|}{anti-vaccine} \\
\hline
 & \textbf{Analytic} & \textbf{Clout} & \textbf{Authentic} & \textbf{Tone} \\
\hline
\textit{C} & 51.25	& 1.07 &	88.50 &	97.19 \\
\hline
\textit{G} & 52.71 & 	2.31 &	43.37 &	1.00 \\
\hline
\end{tabularx}
\end{table}


\subsubsection{Opinion Expression Actions}
To explore how one authentic representation influences the choice of an opinion expression action, we ran an analysis for the pro-vaccine group which was presented with the malware-manipulated post. When controlling for age, a statistically significant difference was found only for the "start following the Twitter account" option - $\chi^2 = 13.012$, $p = 0.023$. The participants in the [43 and older] bracket were less likely to start following a Twitter account that presents an anti-vaccine climate of opinion compared to the other age brackets. When controlling for level of education, a statistically significant difference was found for the "retweet the post" option - $\chi^2 = 12.723$, $p = 0.005$, and for the "start following the Twitter account" option - $\chi^2 = 7.856$, $p = 0.049$. The participants with a high school degree were more likely to retweet the Twitter post and start following a Twitter account that presents an anti-vaccine climate of opinion compared to the participants with higher level of education. No statistical significance was found for the relationships between gender identity and the opinion expression actions on Twitter. These results reveal that a malicious actor could aim to introduce misperception based on the age and the level of education of a targeted Twitter user to push them away from Twitter accounts with otherwise similar stances to their own in regards to the vaccines' effectiveness. 

\section{Discussion}
This study tested the possibility of engineering the spiral-of-silence effect on Twitter by employing a malware-induced misperception about vaccines' risk of causing autism. The results, confirming the dominant pro-vaccine sentiment on Twitter \cite{Blankenship}, show that misperceiving a majority of opposing opinion climate is sufficient to "nudge" a rather certain pro-vaccine user to refrain from tweeting their personal opinion in response. The malware is able to "nudge" the pro-vaccine users to refrain from any endorsement action as liking, retweeting, blocking, or following an account. We extended the analysis to see if the user authentic representation could be used for profiling and targeting with the malware. The results show that only user's education level, and not age and gender, predicts their susceptibility to malware-manipulated misperceptions. Age matters only when a user decides to endorse a tweet, with the 43 years and older ones being less likely to do so. 

We found, as evidenced in the previous spiral-of-silence studies, that the misperception-inducing malware could provoke self-response for the pro-vaccine ones with "hard core" attitudes towards the controversial issue debated \cite{Newman}, \cite{Gearhart2}. Their tweets conveyed a stark, to-the-point, and often negative sentiment as a way of disrupting the spiral-of-silence in a vaccines debate on Twitter. We also found that the female anti-vaccine participants with "hard core" attitudes could be nudged to disrupt the spiral-of-silence if the malware was used to make an originally anti-vaccine tweet appear as it is pro-vaccine (opposite of our study). We emulated a scenario where participants, instead of tweeting their own response, were given a choice to respond using a ready-made suggested response. Our findings indicated that the recommended set of readily available tweets were a much more attractive strategy for the majority of the participants, regardless of their attitude and perceived issue importance of vaccines' link to autism.

\subsection {Defenses and Prevention}
The first line of defense against misperception-inducing malware would require elimination of any suspicious extensions in the Chrome store that require permissions to control how the browser text is presented to a user (alternatively a suspicious third-party Twitter app from an App Store). An example of defense, along the lines of malicious software detection, would be using trusted browsers to detect JavaScript executions that are rearranging words and sentences in the textual portion of an HTML document \cite{Kohlbrenner}. Another example is Chrome's Manifest v3 API that is designed to eliminate extensions exhibiting suspicious behaviour in content manipulation \cite{Google}. Content-level signing might not help in these regards because the content manipulation happens after the content integrity check in the sequence of HTML reception and display. Even with these cautions, a malicious actor may find a way to deploy the malware on a target's browser (for example, an insider threat). As with any social engineering tactic, awareness of trolling and social bots campaigns on social media is an advantage to the users and a second line of defense. We believe an analysis of the Twitter content and metrics in the broader context of the issue at stake could potently raise suspicion about the validity of the perceived majority and minority opinions \cite{Levine}. 

\subsection{Vaccines Chatbot}
The aforementioned awareness method might be costly and inconvenient for most of the Twitter users who do want to participate in the vaccines debate on Twitter. For this purpose, we offered an option for the participants to use a ready-made response as a way of disrupting the spiral-of-silence effect. The overwhelmingly positive results inspired us to develop an assistive chatbot that could be useful in countering the effect of the malware but also any actual troll or a social bot. We got this idea from the custom chatbot Facebook created for its employees to use when going home for Thanksgiving \cite{Frenkel}. The purpose of this chatbot is to offer a ready made-response to the employees and help them break out of the spiral-of-silence in case their families created a hostile climate of opinion towards Facebook, i.e. confronted them with questions about the perceived inadequacies in Facebook's dealing with issues like election meddling and data surveillance. 

For our chatbot, we took a "hostile opinion climate" tweet that the user would be tempted to respond to as an input in a machine learning algorithm that suggested the best tweet response from hundreds of existing twitter responses as an output. To select a response, the algorithm calculated the response that was the closest to the input based on the Euclidean distance, with an additional random number added (to help prevent ties, and to help alternate between responses), measured using features of whether a keyword from a given anti-vaccine or pro-vaccine sentiment exists in the tweets \cite{Blankenship}. For our implementation we used a corpus of originally pro-vaccine tweets by the Center for Disease Control (CDC) and applied the misperception-inducing logic to to fit an anti-vaccine agenda to meet the participant preferences in our study. Although we presented the outputs of the chatbot as suggested opinion expression strategies, one is able to  implement this chatbot in a single Excel workbook, allowing a them to copy and paste a "hostile opinion climate" tweet into the Excel workbook and the workbook would suggest a response to the user. This is a low-cost and easy implementation because does not require any programming knowledge or experience Excel is commonly available tool (the implementation logic is available on demand). 


\subsection{Limitations and Future Work}
Though the results of this study suggest that covertly induced misperception with a malware could engineer the spiral-of-silence effect in a Twitter debate on vaccines, caution is warranted when interpreting them. The use of a controlled tweet allowed us to capture the initial reaction of the participants, but a tweet with multiple comments, different content and metrics, or multiple tweets in a series could have a different effect on the opinion expression behavior. The polarizing topic, the attitude certainty, and the issue importance affected the distribution of our sample, which in turn also limits the generalization of the results. We didn't explicitly ask whether participants will express their opinion if anonymity is granted, but further research should test the malware effect under conditions of anonymity. The malware was tested in its extension variant, but there are many people that access Twitter through smartphone applications or multiple interfaces in the same time. There is a possibility that the same results might not be obtained because smartphone applications provide a different set of interaction affordances and multiple interfaces contribute to repetitive exposure to the same information which can lead to changes in perceptions about the issue importance and one's attitude certainty. The outcomes of the study may be different if user awareness about this malware is raised, as it is usually the case with similar manipulation tactics. The malware tactic might be hard to scale up quickly to a large Twitter population like the trolling or social bots do, but that is what makes the malware compelling to a malicious actor interested in micro-targeting users.

For our next research steps we plan to replicate and extend the current study with Reddit to explore whether the affordances of this particular social media platform affects the malware effectiveness. Our plan is also to cover other polarizing topics popular on Twitter, for example a pandemic virus or responses to the president's tweets. We will work on diversifying our future samples and control for other demographic and cultural factors to get a more nuanced idea of how a spiral-of-silence effect, engineered  by a covert malware, might unfold in the future for a purpose of a covert, low-intensity political propaganda. Towards a more robust test of the malware, the future research will investigate whether a different packaging, e.g. a third-party smartphone twitter application, could amplify or attenuate the misperception-inducing potential of the malware. Another line of research will continue to explore machine learning mechanisms for automated decision making on what type of content rearrangement is the best suited for a particular polarizing issue, target, or a social media platform. Our objective in future research is not to perpetuate any deviant cybersecurity behaviour, but rather the contrary. We are strongly dedicated to investigating any facet of this opinion manipulation method to be able to eradicate it with both technological and societal prevention mechanisms. 

\subsection {Ethical Implications}  
The ethical implications of our study  are the same as those related to publishing any vulnerability: the value of publicly sharing a proof-of-concept exploit with knowledgeable researchers outweighs the opportunity that potential attackers may benefit from the publication. If this paper introduces a viable attack in the social media ecosystem due to its simplistic nature, we believe that this might be merely a confirmation of similar exploits, independently developed and deployed by well-resourced malicious actors. The study itself tests the plausibility of a locally developed browser extension (not publicly available on the Chrome store). In the context of a real-life malware, a responsible disclosure would entail contacting Google, the developers of Chrome, and working with them through the details of the malware extension.  

\section{Conclusion}
In this work, we introduced a misperception-inducing malware as a means of covert opinion manipulation of polarized discourse on Twitter. We tested it with 315 participants and showed that the malware attack has the potential to silence users in both expressing their opinion or taking any opinion endorsement actions. Our main contribution is the evidence that the spiral-of-silence effect can be induced on demand only with a piece of seemingly benign JavaScript (or other software) code and without fabricating any tweets or using bot accounts. We hope our results inform the security community about the implications of having an alternative method for social media influence, at least in a micro-targeted variant. 



\bibliographystyle{ACM-Reference-Format}
\bibliography{sample-base}


\begin{thebibliography}{67}


\ifx \showCODEN    \undefined \def \showCODEN     #1{\unskip}     \fi
\ifx \showDOI      \undefined \def \showDOI       #1{#1}\fi
\ifx \showISBNx    \undefined \def \showISBNx     #1{\unskip}     \fi
\ifx \showISBNxiii \undefined \def \showISBNxiii  #1{\unskip}     \fi
\ifx \showISSN     \undefined \def \showISSN      #1{\unskip}     \fi
\ifx \showLCCN     \undefined \def \showLCCN      #1{\unskip}     \fi
\ifx \shownote     \undefined \def \shownote      #1{#1}          \fi
\ifx \showarticletitle \undefined \def \showarticletitle #1{#1}   \fi
\ifx \showURL      \undefined \def \showURL       {\relax}        \fi
\providecommand\bibfield[2]{#2}
\providecommand\bibinfo[2]{#2}
\providecommand\natexlab[1]{#1}
\providecommand\showeprint[2][]{arXiv:#2}

\bibitem[\protect\citeauthoryear{??}{Chu}{2017}]%
        {Chun}
 \bibinfo{year}{2017}\natexlab{}.
\newblock \showarticletitle{When does individuals' willingness to speak out
  increase on social media? Perceived social support and perceived
  power/control}.
\newblock \bibinfo{journal}{\emph{Computers in Human Behavior}}
  \bibinfo{volume}{74} (\bibinfo{year}{2017}), \bibinfo{pages}{120--129}.
\newblock
\showISBNx{0747-5632}
\urldef\tempurl%
\url{https://doi.org/10.1016/j.chb.2017.04.010}
\showDOI{\tempurl}


\bibitem[\protect\citeauthoryear{{Addawood}}{{Addawood}}{2018}]%
        {Addawood}
\bibfield{author}{\bibinfo{person}{A. {Addawood}}.}
  \bibinfo{year}{2018}\natexlab{}.
\newblock \showarticletitle{Usage of Scientific References in MMR Vaccination
  Debates on Twitter}. In \bibinfo{booktitle}{\emph{2018 IEEE/ACM International
  Conference on Advances in Social Networks Analysis and Mining (ASONAM)}}.
  \bibinfo{pages}{971--979}.
\newblock
\showISSN{2473-9928}
\urldef\tempurl%
\url{https://doi.org/10.1109/ASONAM.2018.8508385}
\showDOI{\tempurl}


\bibitem[\protect\citeauthoryear{Badawy, Lerman, and Ferrara}{Badawy
  et~al\mbox{.}}{2019}]%
        {Badawy}
\bibfield{author}{\bibinfo{person}{Adam Badawy}, \bibinfo{person}{Kristina
  Lerman}, {and} \bibinfo{person}{Emilio Ferrara}.}
  \bibinfo{year}{2019}\natexlab{}.
\newblock \showarticletitle{Who Falls for Online Political Manipulation?}. In
  \bibinfo{booktitle}{\emph{Companion Proceedings of The 2019 World Wide Web
  Conference}} (San Francisco, USA) \emph{(\bibinfo{series}{WWW '19})}.
  \bibinfo{publisher}{ACM}, \bibinfo{address}{New York, NY, USA},
  \bibinfo{pages}{162--168}.
\newblock
\showISBNx{978-1-4503-6675-5}
\urldef\tempurl%
\url{https://doi.org/10.1145/3308560.3316494}
\showDOI{\tempurl}


\bibitem[\protect\citeauthoryear{{Bello} and {Heckel}}{{Bello} and
  {Heckel}}{2019}]%
        {Bello}
\bibfield{author}{\bibinfo{person}{B.~S. {Bello}} {and} \bibinfo{person}{R.
  {Heckel}}.} \bibinfo{year}{2019}\natexlab{}.
\newblock \showarticletitle{Analyzing the Behaviour of Twitter Bots in Post
  Brexit Politics}. In \bibinfo{booktitle}{\emph{2019 Sixth International
  Conference on Social Networks Analysis, Management and Security (SNAMS)}}.
  \bibinfo{pages}{61--66}.
\newblock
\showISSN{null}
\urldef\tempurl%
\url{https://doi.org/10.1109/SNAMS.2019.8931874}
\showDOI{\tempurl}


\bibitem[\protect\citeauthoryear{Bello-Orgaz, Hernandez-Castro, and
  Camacho}{Bello-Orgaz et~al\mbox{.}}{2017}]%
        {Orgaz}
\bibfield{author}{\bibinfo{person}{Gema Bello-Orgaz}, \bibinfo{person}{Julio
  Hernandez-Castro}, {and} \bibinfo{person}{David Camacho}.}
  \bibinfo{year}{2017}\natexlab{}.
\newblock \showarticletitle{Detecting discussion communities on vaccination in
  twitter}.
\newblock \bibinfo{journal}{\emph{Future Generation Computer Systems}}
  \bibinfo{volume}{66} (\bibinfo{year}{2017}), \bibinfo{pages}{125 -- 136}.
\newblock
\showISSN{0167-739X}
\urldef\tempurl%
\url{https://doi.org/10.1016/j.future.2016.06.032}
\showDOI{\tempurl}


\bibitem[\protect\citeauthoryear{Benkler, Faris, and Roberts}{Benkler
  et~al\mbox{.}}{2018}]%
        {Benkler}
\bibfield{author}{\bibinfo{person}{Yochai Benkler}, \bibinfo{person}{Robert
  Faris}, {and} \bibinfo{person}{Hal Roberts}.}
  \bibinfo{year}{2018}\natexlab{}.
\newblock \bibinfo{booktitle}{\emph{Network Propaganda: Manipulation,
  Disinformation, and Radicalization in American Politics}}.
\newblock \bibinfo{publisher}{Oxford University Press},
  \bibinfo{address}{Oxford, UK}.
\newblock


\bibitem[\protect\citeauthoryear{Blankenship, Goff, Yin, Tse, Fu, Liang,
  Saroha, and Fung}{Blankenship et~al\mbox{.}}{2018}]%
        {Blankenship}
\bibfield{author}{\bibinfo{person}{Elizabeth~B Blankenship},
  \bibinfo{person}{Mary~Elizabeth Goff}, \bibinfo{person}{Jinging Yin},
  \bibinfo{person}{Zion Tsz~Ho Tse}, \bibinfo{person}{King-Wa Fu},
  \bibinfo{person}{Hai Liang}, \bibinfo{person}{Nitin Saroha}, {and}
  \bibinfo{person}{Isaac Chun-Hai Fung}.} \bibinfo{year}{2018}\natexlab{}.
\newblock \showarticletitle{Sentiment, Contents, and Retweets: A Study of Two
  Vaccine-Related Twitter Datasets}.
\newblock \bibinfo{journal}{\emph{The Permanente journal}}
  \bibinfo{volume}{22} (\bibinfo{year}{2018}), \bibinfo{pages}{17--138}.
\newblock


\bibitem[\protect\citeauthoryear{Broniatowski, Jamison, Qi, AlKulaib, Chen,
  Benton, Quinn, and Dredze}{Broniatowski et~al\mbox{.}}{2018}]%
        {Broniatowski}
\bibfield{author}{\bibinfo{person}{David~A. Broniatowski},
  \bibinfo{person}{Amelia~M. Jamison}, \bibinfo{person}{SiHua Qi},
  \bibinfo{person}{Lulwah AlKulaib}, \bibinfo{person}{Tao Chen},
  \bibinfo{person}{Adrian Benton}, \bibinfo{person}{Sandra~C. Quinn}, {and}
  \bibinfo{person}{Mark Dredze}.} \bibinfo{year}{2018}\natexlab{}.
\newblock \showarticletitle{Weaponized Health Communication: Twitter Bots and
  Russian Trolls Amplify the Vaccine Debate}.
\newblock \bibinfo{journal}{\emph{American Journal of Public Health}}
  \bibinfo{volume}{108}, \bibinfo{number}{10} (\bibinfo{year}{2018}),
  \bibinfo{pages}{1378--1384}.
\newblock
\urldef\tempurl%
\url{https://doi.org/10.2105/AJPH.2018.304567}
\showDOI{\tempurl}


\bibitem[\protect\citeauthoryear{Chan}{Chan}{2018}]%
        {Chan}
\bibfield{author}{\bibinfo{person}{Michael Chan}.}
  \bibinfo{year}{2018}\natexlab{}.
\newblock \showarticletitle{Reluctance to Talk About Politics in Face-to-Face
  and Facebook Settings: Examining the Impact of Fear of Isolation, Willingness
  to Self-Censor, and Peer Network Characteristics}.
\newblock \bibinfo{journal}{\emph{Mass Communication and Society}}
  \bibinfo{volume}{21}, \bibinfo{number}{1} (\bibinfo{year}{2018}),
  \bibinfo{pages}{1--23}.
\newblock
\urldef\tempurl%
\url{https://doi.org/10.1080/15205436.2017.1358819}
\showDOI{\tempurl}


\bibitem[\protect\citeauthoryear{Cialdini}{Cialdini}{2007}]%
        {Cialdini}
\bibfield{author}{\bibinfo{person}{Robert~B Cialdini}.}
  \bibinfo{year}{2007}\natexlab{}.
\newblock \bibinfo{booktitle}{\emph{{Influence: the psychology of persuasion;
  Rev. ed.}}}
\newblock \bibinfo{publisher}{Collins}, \bibinfo{address}{New York, NY}.
\newblock
\urldef\tempurl%
\url{http://cds.cern.ch/record/2010777}
\showURL{%
\tempurl}


\bibitem[\protect\citeauthoryear{Cresci, Di~Pietro, Petrocchi, Spognardi, and
  Tesconi}{Cresci et~al\mbox{.}}{2017}]%
        {Cresci}
\bibfield{author}{\bibinfo{person}{Stefano Cresci}, \bibinfo{person}{Roberto
  Di~Pietro}, \bibinfo{person}{Marinella Petrocchi}, \bibinfo{person}{Angelo
  Spognardi}, {and} \bibinfo{person}{Maurizio Tesconi}.}
  \bibinfo{year}{2017}\natexlab{}.
\newblock \showarticletitle{The Paradigm-Shift of Social Spambots: Evidence,
  Theories, and Tools for the Arms Race}. In
  \bibinfo{booktitle}{\emph{Proceedings of the 26th International Conference on
  World Wide Web Companion}} (Perth, Australia) \emph{(\bibinfo{series}{WWW
  ’17 Companion})}. \bibinfo{publisher}{International World Wide Web
  Conferences Steering Committee}, \bibinfo{address}{Republic and Canton of
  Geneva, CHE}, \bibinfo{pages}{963–972}.
\newblock
\showISBNx{9781450349147}
\urldef\tempurl%
\url{https://doi.org/10.1145/3041021.3055135}
\showDOI{\tempurl}


\bibitem[\protect\citeauthoryear{DiResta, Shaffer, Ruppel, Becky, Sullivan,
  David, Matney, Fox, Albright, and Johnson}{DiResta et~al\mbox{.}}{2018}]%
        {DiResta}
\bibfield{author}{\bibinfo{person}{Renee DiResta}, \bibinfo{person}{Kris
  Shaffer}, \bibinfo{person}{Ruppel}, \bibinfo{person}{Becky},
  \bibinfo{person}{Sullivan}, \bibinfo{person}{David}, \bibinfo{person}{Robert
  Matney}, \bibinfo{person}{Ryan Fox}, \bibinfo{person}{Jonathan Albright},
  {and} \bibinfo{person}{Ben Johnson}.} \bibinfo{year}{2018}\natexlab{}.
\newblock \bibinfo{booktitle}{\emph{The Tactics and Tropes of the Internet
  Research Agency}}.
\newblock \bibinfo{type}{Technical Report}. \bibinfo{institution}{New
  Knowledge}.
\newblock


\bibitem[\protect\citeauthoryear{Downey}{Downey}{2018}]%
        {Downey}
\bibfield{author}{\bibinfo{person}{Charlie Downey}.}
  \bibinfo{year}{2018}\natexlab{}.
\newblock \bibinfo{title}{Probably Overthinking It}.
\newblock
\newblock
\urldef\tempurl%
\url{http://allendowney.blogspot.com/2018/02/build-your-own-sotu.html}
\showURL{%
\tempurl}


\bibitem[\protect\citeauthoryear{Duh, Slak~Rupnik, and KoroÅ¡ak}{Duh
  et~al\mbox{.}}{2018}]%
        {Duh}
\bibfield{author}{\bibinfo{person}{Andrej Duh}, \bibinfo{person}{Marjan
  Slak~Rupnik}, {and} \bibinfo{person}{Dean KoroÅ¡ak}.}
  \bibinfo{year}{2018}\natexlab{}.
\newblock \showarticletitle{Collective Behavior of Social Bots Is Encoded in
  Their Temporal Twitter Activity}.
\newblock \bibinfo{journal}{\emph{Big Data}} \bibinfo{volume}{6},
  \bibinfo{number}{2} (\bibinfo{year}{2018}), \bibinfo{pages}{113--123}.
\newblock
\urldef\tempurl%
\url{https://doi.org/10.1089/big.2017.0041}
\showDOI{\tempurl}


\bibitem[\protect\citeauthoryear{Faasse, Chatman, and Martin}{Faasse
  et~al\mbox{.}}{2016}]%
        {Faasse}
\bibfield{author}{\bibinfo{person}{Kate Faasse}, \bibinfo{person}{Casey~J.
  Chatman}, {and} \bibinfo{person}{Leslie~R. Martin}.}
  \bibinfo{year}{2016}\natexlab{}.
\newblock \showarticletitle{A comparison of language use in pro- and
  anti-vaccination comments in response to a high profile Facebook post,}.
\newblock \bibinfo{journal}{\emph{Vaccine}} \bibinfo{volume}{34},
  \bibinfo{number}{47} (\bibinfo{year}{2016}), \bibinfo{pages}{5808 -- 5814}.
\newblock
\showISSN{0264-410X}
\urldef\tempurl%
\url{https://doi.org/10.1016/j.vaccine.2016.09.029}
\showDOI{\tempurl}


\bibitem[\protect\citeauthoryear{Ferreira, Coventry, and Lenzini}{Ferreira
  et~al\mbox{.}}{2015}]%
        {Ferreira}
\bibfield{author}{\bibinfo{person}{Ana Ferreira}, \bibinfo{person}{Lynne
  Coventry}, {and} \bibinfo{person}{Gabriele Lenzini}.}
  \bibinfo{year}{2015}\natexlab{}.
\newblock \showarticletitle{Principles of Persuasion in Social Engineering and
  Their Use in Phishing}. In \bibinfo{booktitle}{\emph{Human Aspects of
  Information Security, Privacy, and Trust}},
  \bibfield{editor}{\bibinfo{person}{Theo Tryfonas} {and}
  \bibinfo{person}{Ioannis Askoxylakis}} (Eds.). \bibinfo{publisher}{Springer
  International Publishing}, \bibinfo{pages}{36--47}.
\newblock
\showISBNx{978-3-319-20376-8}


\bibitem[\protect\citeauthoryear{Frenkel and Isaac}{Frenkel and Isaac}{2019}]%
        {Frenkel}
\bibfield{author}{\bibinfo{person}{Sheera Frenkel} {and} \bibinfo{person}{Mike
  Isaac}.} \bibinfo{year}{2019}\natexlab{}.
\newblock \bibinfo{title}{Facebook Gives Workers a Chatbot to Appease That
  Prying Uncle}.
\newblock
\newblock
\urldef\tempurl%
\url{https://www.nytimes.com/2019/12/02/technology/facebook-chatbot-workers.html?smid=tw-nytimes&smtyp=cur}
\showURL{%
\tempurl}


\bibitem[\protect\citeauthoryear{Garimella, Morales, Gionis, and
  Mathioudakis}{Garimella et~al\mbox{.}}{2018}]%
        {Kiran}
\bibfield{author}{\bibinfo{person}{Kiran Garimella}, \bibinfo{person}{Gianmarco
  De~Francisci Morales}, \bibinfo{person}{Aristides Gionis}, {and}
  \bibinfo{person}{Michael Mathioudakis}.} \bibinfo{year}{2018}\natexlab{}.
\newblock \showarticletitle{Quantifying Controversy on Social Media}.
\newblock \bibinfo{journal}{\emph{Trans. Soc. Comput.}} \bibinfo{volume}{1},
  \bibinfo{number}{1}, Article \bibinfo{articleno}{3} (\bibinfo{date}{Jan.}
  \bibinfo{year}{2018}), \bibinfo{numpages}{27}~pages.
\newblock
\showISSN{2469-7818}
\urldef\tempurl%
\url{https://doi.org/10.1145/3140565}
\showDOI{\tempurl}


\bibitem[\protect\citeauthoryear{Gearhart and Zhang}{Gearhart and
  Zhang}{2013}]%
        {Gearhart}
\bibfield{author}{\bibinfo{person}{Sherice Gearhart} {and}
  \bibinfo{person}{Weiwu Zhang}.} \bibinfo{year}{2013}\natexlab{}.
\newblock \showarticletitle{Gay Bullying and Online Opinion Expression: Testing
  Spiral of Silence in the Social Media Environment}.
\newblock \bibinfo{journal}{\emph{Social Science Computer Review}}
  \bibinfo{volume}{32}, \bibinfo{number}{1} (\bibinfo{date}{2019/09/03}
  \bibinfo{year}{2013}), \bibinfo{pages}{18--36}.
\newblock
\showISBNx{0894-4393}
\urldef\tempurl%
\url{https://doi.org/10.1177/0894439313504261}
\showDOI{\tempurl}


\bibitem[\protect\citeauthoryear{Gearhart and Zhang}{Gearhart and
  Zhang}{2015}]%
        {Gearhart2}
\bibfield{author}{\bibinfo{person}{Sherice Gearhart} {and}
  \bibinfo{person}{Weiwu Zhang}.} \bibinfo{year}{2015}\natexlab{}.
\newblock \showarticletitle{``Was It Something I Said?''``No, It Was Something
  You Posted!''A Study of the Spiral of Silence Theory in Social Media
  Contexts}.
\newblock \bibinfo{journal}{\emph{Cyberpsychology, Behavior, and Social
  Networking}} \bibinfo{volume}{18}, \bibinfo{number}{4}
  (\bibinfo{date}{2019/09/04} \bibinfo{year}{2015}), \bibinfo{pages}{208--213}.
\newblock
\showISBNx{2152-2715}
\urldef\tempurl%
\url{https://doi.org/10.1089/cyber.2014.0443}
\showDOI{\tempurl}


\bibitem[\protect\citeauthoryear{Ghanem, Buscaldi, and Rosso}{Ghanem
  et~al\mbox{.}}{2019}]%
        {Ghanem}
\bibfield{author}{\bibinfo{person}{Bilal Ghanem}, \bibinfo{person}{Davide
  Buscaldi}, {and} \bibinfo{person}{Paolo Rosso}.}
  \bibinfo{year}{2019}\natexlab{}.
\newblock \bibinfo{title}{TexTrolls: Identifying Russian Trolls on Twitter from
  a Textual Perspective}.
\newblock
\newblock
\showeprint[arxiv]{cs.CL/1910.01340}


\bibitem[\protect\citeauthoryear{Google}{Google}{2018}]%
        {Google}
\bibfield{author}{\bibinfo{person}{Google}.} \bibinfo{year}{2018}\natexlab{}.
\newblock \bibinfo{title}{Manifest V3}.
\newblock
\newblock
\urldef\tempurl%
\url{https://docs.google.com/document/d/1nPu6Wy4LWR66EFLeYInl3NzzhHzc-qnk4w4PX-0XMw8/edit}
\showURL{%
\tempurl}


\bibitem[\protect\citeauthoryear{Grosser}{Grosser}{2018}]%
        {Grosser}
\bibfield{author}{\bibinfo{person}{Ben Grosser}.}
  \bibinfo{year}{2018}\natexlab{}.
\newblock \bibinfo{title}{{Twitter Demetricator | benjamin grosser}}.
\newblock
\newblock
\urldef\tempurl%
\url{https://bengrosser.com/projects/twitter-demetricator/}
\showURL{%
\tempurl}


\bibitem[\protect\citeauthoryear{Grover, Kar, Dwivedi, and Janssen}{Grover
  et~al\mbox{.}}{2019}]%
        {Grover}
\bibfield{author}{\bibinfo{person}{Purva Grover}, \bibinfo{person}{Arpan~Kumar
  Kar}, \bibinfo{person}{Yogesh~K. Dwivedi}, {and} \bibinfo{person}{Marijn
  Janssen}.} \bibinfo{year}{2019}\natexlab{}.
\newblock \showarticletitle{Polarization and acculturation in US Election 2016
  outcomes - Can twitter analytics predict changes in voting preferences}.
\newblock \bibinfo{journal}{\emph{Technological Forecasting and Social Change}}
   \bibinfo{volume}{145} (\bibinfo{year}{2019}), \bibinfo{pages}{438 -- 460}.
\newblock
\showISSN{0040-1625}
\urldef\tempurl%
\url{https://doi.org/10.1016/j.techfore.2018.09.009}
\showDOI{\tempurl}


\bibitem[\protect\citeauthoryear{Hampton, Rainie, Lu, Dwyer, Shin, and
  Purcell}{Hampton et~al\mbox{.}}{2014}]%
        {Hampton}
\bibfield{author}{\bibinfo{person}{Keith Hampton}, \bibinfo{person}{Lee
  Rainie}, \bibinfo{person}{Weixu Lu}, \bibinfo{person}{Maria Dwyer},
  \bibinfo{person}{Inyoung Shin}, {and} \bibinfo{person}{Kristen Purcell}.}
  \bibinfo{year}{2014}\natexlab{}.
\newblock \bibinfo{booktitle}{\emph{Social Media and the `Spiral of Silence'}}.
\newblock \bibinfo{type}{Technical Report}. \bibinfo{institution}{Pew Research
  Center}, \bibinfo{address}{Washington DC}.
\newblock


\bibitem[\protect\citeauthoryear{Hardaker}{Hardaker}{2010}]%
        {Hardaker}
\bibfield{author}{\bibinfo{person}{Claire Hardaker}.}
  \bibinfo{year}{2010}\natexlab{}.
\newblock \showarticletitle{Trolling in asynchronous computer-mediated
  communication: From user discussions to academic definitions}.
\newblock \bibinfo{journal}{\emph{Journal of Politeness Research}}
  \bibinfo{volume}{6} (\bibinfo{year}{2010}), \bibinfo{pages}{215--242}.
\newblock
Issue 2.
\urldef\tempurl%
\url{https://doi.org/10.1515/jplr.2010.011}
\showDOI{\tempurl}


\bibitem[\protect\citeauthoryear{Hayes, Glynn, and Shanahan}{Hayes
  et~al\mbox{.}}{2005}]%
        {Hayes}
\bibfield{author}{\bibinfo{person}{Andrew~F. Hayes},
  \bibinfo{person}{Carroll~J. Glynn}, {and} \bibinfo{person}{James Shanahan}.}
  \bibinfo{year}{2005}\natexlab{}.
\newblock \showarticletitle{Willingness to Self-Censor: A Construct and
  Measurement Tool for Public Opinion Research}.
\newblock \bibinfo{journal}{\emph{International Journal of Public Opinion
  Research}} \bibinfo{volume}{17}, \bibinfo{number}{3}
  (\bibinfo{date}{9/5/2019} \bibinfo{year}{2005}), \bibinfo{pages}{298--323}.
\newblock
\showISBNx{1471-6909}
\urldef\tempurl%
\url{https://doi.org/10.1093/ijpor/edh073}
\showDOI{\tempurl}


\bibitem[\protect\citeauthoryear{Hoffmann and Lutz}{Hoffmann and Lutz}{2017}]%
        {Hoffmann}
\bibfield{author}{\bibinfo{person}{Christian~Pieter Hoffmann} {and}
  \bibinfo{person}{Christoph Lutz}.} \bibinfo{year}{2017}\natexlab{}.
\newblock \showarticletitle{Spiral of Silence 2.0: Political Self-Censorship
  among Young Facebook Users}. In \bibinfo{booktitle}{\emph{Proceedings of the
  8th International Conference on Social Media \& Society}} (Toronto, ON,
  Canada). \bibinfo{publisher}{Association for Computing Machinery},
  \bibinfo{address}{New York, NY, USA}, Article \bibinfo{articleno}{10},
  \bibinfo{numpages}{12}~pages.
\newblock
\showISBNx{9781450348478}
\urldef\tempurl%
\url{https://doi.org/10.1145/3097286.3097296}
\showDOI{\tempurl}


\bibitem[\protect\citeauthoryear{Jamison, Broniatowski, and Quinn}{Jamison
  et~al\mbox{.}}{2019}]%
        {Jamison}
\bibfield{author}{\bibinfo{person}{Amelia~M. Jamison},
  \bibinfo{person}{David~A. Broniatowski}, {and} \bibinfo{person}{Sandra~Crouse
  Quinn}.} \bibinfo{year}{2019}\natexlab{}.
\newblock \showarticletitle{Malicious Actors on Twitter: A Guide for Public
  Health Researchers}.
\newblock \bibinfo{journal}{\emph{American Journal of Public Health}}
  \bibinfo{volume}{109}, \bibinfo{number}{5} (\bibinfo{year}{2019}),
  \bibinfo{pages}{688--692}.
\newblock
\urldef\tempurl%
\url{https://doi.org/10.2105/AJPH.2019.304969}
\showDOI{\tempurl}


\bibitem[\protect\citeauthoryear{Jang, Song, Chung, Wang, and Lee}{Jang
  et~al\mbox{.}}{2014}]%
        {Jang}
\bibfield{author}{\bibinfo{person}{Yeongjin Jang}, \bibinfo{person}{Chengyu
  Song}, \bibinfo{person}{Simon~P. Chung}, \bibinfo{person}{Tielei Wang}, {and}
  \bibinfo{person}{Wenke Lee}.} \bibinfo{year}{2014}\natexlab{}.
\newblock \showarticletitle{{A11Y Attacks: Exploiting Accessibility in
  Operating Systems}}. In \bibinfo{booktitle}{\emph{Proceedings of the 2014 ACM
  SIGSAC Conference on Computer and Communications Security}} (Scottsdale,
  Arizona, USA) \emph{(\bibinfo{series}{CCS '14})}. \bibinfo{publisher}{ACM},
  \bibinfo{address}{New York, NY, USA}, \bibinfo{pages}{103--115}.
\newblock
\showISBNx{978-1-4503-2957-6}
\urldef\tempurl%
\url{https://doi.org/10.1145/2660267.2660295}
\showDOI{\tempurl}


\bibitem[\protect\citeauthoryear{Kata}{Kata}{2012}]%
        {Kata}
\bibfield{author}{\bibinfo{person}{Anna Kata}.}
  \bibinfo{year}{2012}\natexlab{}.
\newblock \showarticletitle{Anti-vaccine activists, Web 2.0, and the postmodern
  paradigm - an overview of tactics and tropes used online by the
  anti-vaccination movement}.
\newblock \bibinfo{journal}{\emph{Vaccine}} \bibinfo{volume}{30},
  \bibinfo{number}{25} (\bibinfo{year}{2012}), \bibinfo{pages}{3778 -- 3789}.
\newblock
\showISSN{0264-410X}
\urldef\tempurl%
\url{https://doi.org/10.1016/j.vaccine.2011.11.112}
\showDOI{\tempurl}
\newblock
\shownote{Special Issue: The Role of Internet Use in Vaccination Decisions.}


\bibitem[\protect\citeauthoryear{Kim}{Kim}{2016}]%
        {Kim}
\bibfield{author}{\bibinfo{person}{Mihee Kim}.}
  \bibinfo{year}{2016}\natexlab{}.
\newblock \showarticletitle{Facebook's Spiral of Silence and Participation: The
  Role of Political Expression on Facebook and Partisan Strength in Political
  Participation}.
\newblock \bibinfo{journal}{\emph{Cyberpsychology, Behavior, and Social
  Networking}} \bibinfo{volume}{19}, \bibinfo{number}{12}
  (\bibinfo{date}{2019/09/08} \bibinfo{year}{2016}), \bibinfo{pages}{696--702}.
\newblock
\showISBNx{2152-2715}
\urldef\tempurl%
\url{https://doi.org/10.1089/cyber.2016.0137}
\showDOI{\tempurl}


\bibitem[\protect\citeauthoryear{Kirman, Lineham, and Lawson}{Kirman
  et~al\mbox{.}}{2012}]%
        {Kirman}
\bibfield{author}{\bibinfo{person}{Ben Kirman}, \bibinfo{person}{Conor
  Lineham}, {and} \bibinfo{person}{Shaun Lawson}.}
  \bibinfo{year}{2012}\natexlab{}.
\newblock \showarticletitle{Exploring Mischief and Mayhem in Social Computing
  or: How We Learned to Stop Worrying and Love the Trolls}. In
  \bibinfo{booktitle}{\emph{CHI '12 Extended Abstracts on Human Factors in
  Computing Systems}} (Austin, Texas, USA) \emph{(\bibinfo{series}{CHI EA
  '12})}. \bibinfo{publisher}{ACM}, \bibinfo{address}{New York, NY, USA},
  \bibinfo{pages}{121--130}.
\newblock
\showISBNx{978-1-4503-1016-1}
\urldef\tempurl%
\url{https://doi.org/10.1145/2212776.2212790}
\showDOI{\tempurl}


\bibitem[\protect\citeauthoryear{Kohlbrenner and Shacham}{Kohlbrenner and
  Shacham}{2016}]%
        {Kohlbrenner}
\bibfield{author}{\bibinfo{person}{David Kohlbrenner} {and}
  \bibinfo{person}{Hovav Shacham}.} \bibinfo{year}{2016}\natexlab{}.
\newblock \showarticletitle{Trusted Browsers for Uncertain Times}. In
  \bibinfo{booktitle}{\emph{25th {USENIX} Security Symposium ({USENIX} Security
  16)}}. \bibinfo{publisher}{{USENIX} Association}, \bibinfo{address}{Austin,
  TX}, \bibinfo{pages}{463--480}.
\newblock
\showISBNx{978-1-931971-32-4}
\urldef\tempurl%
\url{https://www.usenix.org/conference/usenixsecurity16/technical-sessions/presentation/kohlbrenner}
\showURL{%
\tempurl}


\bibitem[\protect\citeauthoryear{Koutra, Bennett, and Horvitz}{Koutra
  et~al\mbox{.}}{2015}]%
        {Danai}
\bibfield{author}{\bibinfo{person}{Danai Koutra}, \bibinfo{person}{Paul~N.
  Bennett}, {and} \bibinfo{person}{Eric Horvitz}.}
  \bibinfo{year}{2015}\natexlab{}.
\newblock \showarticletitle{Events and Controversies: Influences of a Shocking
  News Event on Information Seeking}. In \bibinfo{booktitle}{\emph{Proceedings
  of the 24th International Conference on World Wide Web}} (Florence, Italy)
  \emph{(\bibinfo{series}{WWW ’15})}. \bibinfo{publisher}{International World
  Wide Web Conferences Steering Committee}, \bibinfo{address}{Republic and
  Canton of Geneva, CHE}, \bibinfo{pages}{614–624}.
\newblock
\showISBNx{9781450334693}
\urldef\tempurl%
\url{https://doi.org/10.1145/2736277.2741099}
\showDOI{\tempurl}


\bibitem[\protect\citeauthoryear{Kushin, Yamamoto, and Dalisay}{Kushin
  et~al\mbox{.}}{2019}]%
        {Kushin}
\bibfield{author}{\bibinfo{person}{Matthew~J. Kushin},
  \bibinfo{person}{Masahiro Yamamoto}, {and} \bibinfo{person}{Francis
  Dalisay}.} \bibinfo{year}{2019}\natexlab{}.
\newblock \showarticletitle{Societal Majority, Facebook, and the Spiral of
  Silence in the 2016 US Presidential Election}.
\newblock \bibinfo{journal}{\emph{Social Media + Society}}  \bibinfo{volume}{5}
  (\bibinfo{date}{2020/01/01} \bibinfo{year}{2019}),
  \bibinfo{pages}{2056305119855139}.
\newblock
\showISBNx{2056-3051}
\urldef\tempurl%
\url{https://doi.org/10.1177/2056305119855139}
\showDOI{\tempurl}


\bibitem[\protect\citeauthoryear{Lee and Kim}{Lee and Kim}{2014a}]%
        {Lee}
\bibfield{author}{\bibinfo{person}{Na~Yeon Lee} {and} \bibinfo{person}{Yonghwan
  Kim}.} \bibinfo{year}{2014}\natexlab{a}.
\newblock \showarticletitle{The spiral of silence and journalists'
  outspokenness on Twitter}.
\newblock \bibinfo{journal}{\emph{Asian Journal of Communication}}
  \bibinfo{volume}{24}, \bibinfo{number}{3} (\bibinfo{date}{05}
  \bibinfo{year}{2014}), \bibinfo{pages}{262--278}.
\newblock
\showISBNx{0129-2986}
\urldef\tempurl%
\url{https://doi.org/10.1080/01292986.2014.885536}
\showDOI{\tempurl}


\bibitem[\protect\citeauthoryear{Lee and Kim}{Lee and Kim}{2014b}]%
        {Lee1}
\bibfield{author}{\bibinfo{person}{Na~Yeon Lee} {and} \bibinfo{person}{Yonghwan
  Kim}.} \bibinfo{year}{2014}\natexlab{b}.
\newblock \showarticletitle{The spiral of silence and journalists'
  outspokenness on Twitter}.
\newblock \bibinfo{journal}{\emph{Asian Journal of Communication}}
  \bibinfo{volume}{24}, \bibinfo{number}{3} (\bibinfo{year}{2014}),
  \bibinfo{pages}{262--278}.
\newblock
\urldef\tempurl%
\url{https://doi.org/10.1080/01292986.2014.885536}
\showDOI{\tempurl}


\bibitem[\protect\citeauthoryear{Levine}{Levine}{2014}]%
        {Levine}
\bibfield{author}{\bibinfo{person}{Timothy~R Levine}.}
  \bibinfo{year}{2014}\natexlab{}.
\newblock \bibinfo{title}{{Truth-Default} Theory ({TDT})}.
\newblock , \bibinfo{numpages}{378--392}~pages.
\newblock


\bibitem[\protect\citeauthoryear{Lin and Salwen}{Lin and Salwen}{1997}]%
        {LinSalawen}
\bibfield{author}{\bibinfo{person}{Carolyn~A. Lin} {and}
  \bibinfo{person}{Michael~B. Salwen}.} \bibinfo{year}{1997}\natexlab{}.
\newblock \showarticletitle{Predicting the spiral of silence on a controversial
  public issue}.
\newblock \bibinfo{journal}{\emph{Howard Journal of Communications}}
  \bibinfo{volume}{8}, \bibinfo{number}{1} (\bibinfo{date}{01}
  \bibinfo{year}{1997}), \bibinfo{pages}{129--141}.
\newblock
\showISBNx{1064-6175}
\urldef\tempurl%
\url{https://doi.org/10.1080/10646179709361747}
\showDOI{\tempurl}


\bibitem[\protect\citeauthoryear{Liu, Rui, and Cui}{Liu et~al\mbox{.}}{2017}]%
        {Liu}
\bibfield{author}{\bibinfo{person}{Yu Liu}, \bibinfo{person}{Jian~Raymond Rui},
  {and} \bibinfo{person}{Xi Cui}.} \bibinfo{year}{2017}\natexlab{}.
\newblock \showarticletitle{Are people willing to share their political
  opinions on Facebook? Exploring roles of self-presentational concern in
  spiral of silence}.
\newblock \bibinfo{journal}{\emph{Computers in Human Behavior}}
  \bibinfo{volume}{76} (\bibinfo{year}{2017}), \bibinfo{pages}{294--302}.
\newblock
\showISBNx{0747-5632}
\urldef\tempurl%
\url{http://www.sciencedirect.com/science/article/pii/S074756321730451X}
\showURL{%
\tempurl}


\bibitem[\protect\citeauthoryear{Llewellyn, Cram, Favero, and Hill}{Llewellyn
  et~al\mbox{.}}{2018}]%
        {Llewellyn}
\bibfield{author}{\bibinfo{person}{Clare Llewellyn}, \bibinfo{person}{Laura
  Cram}, \bibinfo{person}{Adrian Favero}, {and} \bibinfo{person}{Robin~L.
  Hill}.} \bibinfo{year}{2018}\natexlab{}.
\newblock \showarticletitle{Russian Troll Hunting in a Brexit Twitter Archive}.
  In \bibinfo{booktitle}{\emph{Proceedings of the 18th ACM/IEEE on Joint
  Conference on Digital Libraries}} (Fort Worth, Texas, USA)
  \emph{(\bibinfo{series}{JCDL ’18})}. \bibinfo{publisher}{Association for
  Computing Machinery}, \bibinfo{address}{New York, NY, USA},
  \bibinfo{pages}{361–362}.
\newblock
\showISBNx{9781450351782}
\urldef\tempurl%
\url{https://doi.org/10.1145/3197026.3203876}
\showDOI{\tempurl}


\bibitem[\protect\citeauthoryear{Love, Himelboim, Holton, and Stewart}{Love
  et~al\mbox{.}}{2013}]%
        {Love}
\bibfield{author}{\bibinfo{person}{Brad Love}, \bibinfo{person}{Itai
  Himelboim}, \bibinfo{person}{Avery Holton}, {and} \bibinfo{person}{Kristin
  Stewart}.} \bibinfo{year}{2013}\natexlab{}.
\newblock \showarticletitle{Twitter as a source of vaccination information:
  Content drivers and what they are saying}.
\newblock \bibinfo{journal}{\emph{American Journal of Infection Control}}
  \bibinfo{volume}{41}, \bibinfo{number}{6} (\bibinfo{year}{2013}),
  \bibinfo{pages}{568 -- 570}.
\newblock
\showISSN{0196-6553}
\urldef\tempurl%
\url{https://doi.org/10.1016/j.ajic.2012.10.016}
\showDOI{\tempurl}


\bibitem[\protect\citeauthoryear{Luceri, Deb, Badawy, and Ferrara}{Luceri
  et~al\mbox{.}}{2019}]%
        {Luceri}
\bibfield{author}{\bibinfo{person}{Luca Luceri}, \bibinfo{person}{Ashok Deb},
  \bibinfo{person}{Adam Badawy}, {and} \bibinfo{person}{Emilio Ferrara}.}
  \bibinfo{year}{2019}\natexlab{}.
\newblock \showarticletitle{Red Bots Do It Better:Comparative Analysis of
  Social Bot Partisan Behavior}. In \bibinfo{booktitle}{\emph{Companion
  Proceedings of The 2019 World Wide Web Conference}} (San Francisco, USA)
  \emph{(\bibinfo{series}{WWW ’19})}. \bibinfo{publisher}{Association for
  Computing Machinery}, \bibinfo{address}{New York, NY, USA},
  \bibinfo{pages}{1007–1012}.
\newblock
\showISBNx{9781450366755}
\urldef\tempurl%
\url{https://doi.org/10.1145/3308560.3316735}
\showDOI{\tempurl}


\bibitem[\protect\citeauthoryear{Matthes, Knoll, and von Sikorski}{Matthes
  et~al\mbox{.}}{2018}]%
        {Matthes}
\bibfield{author}{\bibinfo{person}{J{\"o}rg Matthes}, \bibinfo{person}{Johannes
  Knoll}, {and} \bibinfo{person}{Christian von Sikorski}.}
  \bibinfo{year}{2018}\natexlab{}.
\newblock \showarticletitle{The ``Spiral of Silence'' Revisited: A
  Meta-Analysis on the Relationship Between Perceptions of Opinion Support and
  Political Opinion Expression}.
\newblock \bibinfo{journal}{\emph{Communication Research}}
  \bibinfo{volume}{45}, \bibinfo{number}{1} (\bibinfo{year}{2018}),
  \bibinfo{pages}{3--33}.
\newblock
\urldef\tempurl%
\url{https://doi.org/10.1177/0093650217745429}
\showDOI{\tempurl}
\showeprint{https://doi.org/10.1177/0093650217745429}


\bibitem[\protect\citeauthoryear{McKeever, McKeever, Holton, and Li}{McKeever
  et~al\mbox{.}}{2016}]%
        {McKeever}
\bibfield{author}{\bibinfo{person}{Brooke~Weberling McKeever},
  \bibinfo{person}{Robert McKeever}, \bibinfo{person}{Avery~E. Holton}, {and}
  \bibinfo{person}{Jo-Yun Li}.} \bibinfo{year}{2016}\natexlab{}.
\newblock \showarticletitle{Silent Majority: Childhood Vaccinations and
  Antecedents to Communicative Action}.
\newblock \bibinfo{journal}{\emph{Mass Communication and Society}}
  \bibinfo{volume}{19}, \bibinfo{number}{4} (\bibinfo{year}{2016}),
  \bibinfo{pages}{476--498}.
\newblock
\urldef\tempurl%
\url{https://doi.org/10.1080/15205436.2016.1148172}
\showDOI{\tempurl}


\bibitem[\protect\citeauthoryear{Mitra, Counts, and Pennebaker}{Mitra
  et~al\mbox{.}}{2016}]%
        {Mitra}
\bibfield{author}{\bibinfo{person}{Tanushree Mitra}, \bibinfo{person}{Scott
  Counts}, {and} \bibinfo{person}{James~W Pennebaker}.}
  \bibinfo{year}{2016}\natexlab{}.
\newblock \showarticletitle{Understanding anti-vaccination attitudes in social
  media}. In \bibinfo{booktitle}{\emph{Tenth International AAAI Conference on
  Web and Social Media}}.
\newblock


\bibitem[\protect\citeauthoryear{Morgan}{Morgan}{2019}]%
        {Morgan}
\bibfield{author}{\bibinfo{person}{Callie~Jessica Morgan}.}
  \bibinfo{year}{2019}\natexlab{}.
\newblock \emph{\bibinfo{title}{The Silencing Power of Algorithms: How the
  Facebook News Feed Algorithm Manipulates Users: Perceptions of Opinion
  Climates}}.
\newblock \bibinfo{thesistype}{Ph.D. Dissertation}. \bibinfo{school}{Portland
  State University}.
\newblock


\bibitem[\protect\citeauthoryear{Nekmat and Gonzenbach}{Nekmat and
  Gonzenbach}{2013}]%
        {Nekmat}
\bibfield{author}{\bibinfo{person}{Elmie Nekmat} {and}
  \bibinfo{person}{William~J. Gonzenbach}.} \bibinfo{year}{2013}\natexlab{}.
\newblock \showarticletitle{Multiple Opinion Climates in Online Forums: Role of
  Website Source Reference and Within-Forum Opinion Congruency}.
\newblock \bibinfo{journal}{\emph{Journalism \& Mass Communication Quarterly}}
  \bibinfo{volume}{90}, \bibinfo{number}{4} (\bibinfo{year}{2013}),
  \bibinfo{pages}{736--756}.
\newblock
\urldef\tempurl%
\url{https://doi.org/10.1177/1077699013503162}
\showDOI{\tempurl}


\bibitem[\protect\citeauthoryear{Newman}{Newman}{2018}]%
        {Newman}
\bibfield{author}{\bibinfo{person}{Lily~Hay Newman}.}
  \bibinfo{year}{2018}\natexlab{}.
\newblock \bibinfo{title}{{Chrome Extension Malware Has Evolved}}.
\newblock
\newblock
\urldef\tempurl%
\url{https://www.wired.com/story/chrome-extension-malware/}
\showURL{%
\tempurl}


\bibitem[\protect\citeauthoryear{Noelle-Neumann}{Noelle-Neumann}{1993}]%
        {Noelle-Neumann}
\bibfield{author}{\bibinfo{person}{Elisabeth Noelle-Neumann}.}
  \bibinfo{year}{1993}\natexlab{}.
\newblock \bibinfo{booktitle}{\emph{The Spiral of Silence - Public Opinion: Our
  Social Skin} (\bibinfo{edition}{2nd} ed.)}.
\newblock \bibinfo{publisher}{The University of Chicago Press},
  \bibinfo{address}{Chicago, IL}.
\newblock


\bibitem[\protect\citeauthoryear{Orenstein and Ahmed}{Orenstein and
  Ahmed}{2017}]%
        {Orenstein}
\bibfield{author}{\bibinfo{person}{Walter~A Orenstein} {and}
  \bibinfo{person}{Rafi Ahmed}.} \bibinfo{year}{2017}\natexlab{}.
\newblock \showarticletitle{Simply put: Vaccination saves lives}.
\newblock \bibinfo{journal}{\emph{Proceedings of the National Academy of
  Sciences of the United States of America}} \bibinfo{volume}{114},
  \bibinfo{number}{16} (\bibinfo{date}{04} \bibinfo{year}{2017}),
  \bibinfo{pages}{4031--4033}.
\newblock


\bibitem[\protect\citeauthoryear{Scheufle and Moy}{Scheufle and Moy}{2000}]%
        {Scheufle}
\bibfield{author}{\bibinfo{person}{Dietram~A. Scheufle} {and}
  \bibinfo{person}{Patricia Moy}.} \bibinfo{year}{2000}\natexlab{}.
\newblock \showarticletitle{Twenty-Five Years of the Spiral of Silence: A
  Conceptual Review and Empirical Outlook}.
\newblock \bibinfo{journal}{\emph{International Journal of Public Opinion
  Research}} \bibinfo{volume}{12}, \bibinfo{number}{1}
  (\bibinfo{date}{9/9/2019} \bibinfo{year}{2000}), \bibinfo{pages}{3--28}.
\newblock
\showISBNx{1471-6909}
\urldef\tempurl%
\url{https://doi.org/10.1093/ijpor/12.1.3}
\showDOI{\tempurl}


\bibitem[\protect\citeauthoryear{Seals}{Seals}{2019}]%
        {Seals}
\bibfield{author}{\bibinfo{person}{Tara Seals}.}
  \bibinfo{year}{2019}\natexlab{}.
\newblock \bibinfo{title}{SDKs Misused to Scrape Twitter, Facebook Account
  Info}.
\newblock
\newblock
\urldef\tempurl%
\url{https://threatpost.com/sdks-scrape-personal-info-twitter-facebook/150686/}
\showURL{%
\tempurl}


\bibitem[\protect\citeauthoryear{Spangher, Ranade, Nushi, Fourney, and
  Horvitz}{Spangher et~al\mbox{.}}{2018}]%
        {Spangher}
\bibfield{author}{\bibinfo{person}{Alexander Spangher},
  \bibinfo{person}{Gireeja Ranade}, \bibinfo{person}{Besmira Nushi},
  \bibinfo{person}{Adam Fourney}, {and} \bibinfo{person}{Eric Horvitz}.}
  \bibinfo{year}{2018}\natexlab{}.
\newblock \bibinfo{title}{Analysis of Strategy and Spread of Russia-sponsored
  Content in the US in 2017}.
\newblock
\newblock
\showeprint[arxiv]{cs.SI/1810.10033}


\bibitem[\protect\citeauthoryear{Stewart, Arif, and Starbird}{Stewart
  et~al\mbox{.}}{2018}]%
        {Stewart}
\bibfield{author}{\bibinfo{person}{Leo~G Stewart}, \bibinfo{person}{Ahmer
  Arif}, {and} \bibinfo{person}{Kate Starbird}.}
  \bibinfo{year}{2018}\natexlab{}.
\newblock \showarticletitle{Examining trolls and polarization with a retweet
  network}. In \bibinfo{booktitle}{\emph{Proc. ACM WSDM, workshop on
  misinformation and misbehavior mining on the web. 2018.}}
\newblock


\bibitem[\protect\citeauthoryear{{Straton}, {Jang}, {Ng}, {Vatrapu}, and
  {Mukkamala}}{{Straton} et~al\mbox{.}}{2019}]%
        {Straton}
\bibfield{author}{\bibinfo{person}{N. {Straton}}, \bibinfo{person}{H. {Jang}},
  \bibinfo{person}{R. {Ng}}, \bibinfo{person}{R. {Vatrapu}}, {and}
  \bibinfo{person}{R.~R. {Mukkamala}}.} \bibinfo{year}{2019}\natexlab{}.
\newblock \showarticletitle{Computational modeling of stigmatized behaviour in
  pro-vaccination and anti-vaccination discussions on social media}. In
  \bibinfo{booktitle}{\emph{2019 IEEE International Conference on
  Bioinformatics and Biomedicine (BIBM)}}. \bibinfo{pages}{2673--2681}.
\newblock


\bibitem[\protect\citeauthoryear{Tausczik and Pennebaker}{Tausczik and
  Pennebaker}{2010}]%
        {Pennebaker}
\bibfield{author}{\bibinfo{person}{Yla~R. Tausczik} {and}
  \bibinfo{person}{James~W. Pennebaker}.} \bibinfo{year}{2010}\natexlab{}.
\newblock \showarticletitle{The Psychological Meaning of Words: LIWC and
  Computerized Text Analysis Methods}.
\newblock \bibinfo{journal}{\emph{Journal of Language and Social Psychology}}
  \bibinfo{volume}{29}, \bibinfo{number}{1} (\bibinfo{year}{2010}),
  \bibinfo{pages}{24--54}.
\newblock
\urldef\tempurl%
\url{https://doi.org/10.1177/0261927X09351676}
\showDOI{\tempurl}


\bibitem[\protect\citeauthoryear{Thompson and Lapowski}{Thompson and
  Lapowski}{[n.d.]}]%
        {Thompson}
\bibfield{author}{\bibinfo{person}{Nicholas Thompson} {and}
  \bibinfo{person}{Issie Lapowski}.} \bibinfo{year}{[n.d.]}\natexlab{}.
\newblock \bibinfo{title}{How Russian Trolls Used Meme Warfare to Divide
  America}.
\newblock
\newblock
\urldef\tempurl%
\url{https://www.wired.com/story/russia-ira-propaganda-senate-report/}
\showURL{%
\tempurl}


\bibitem[\protect\citeauthoryear{Tomeny, Vargo, and El-Toukhy}{Tomeny
  et~al\mbox{.}}{2017}]%
        {Tomeny}
\bibfield{author}{\bibinfo{person}{Theodore~S. Tomeny},
  \bibinfo{person}{Christopher~J. Vargo}, {and} \bibinfo{person}{Sherine
  El-Toukhy}.} \bibinfo{year}{2017}\natexlab{}.
\newblock \showarticletitle{Geographic and demographic correlates of
  autism-related anti-vaccine beliefs on Twitter, 2009-15}.
\newblock \bibinfo{journal}{\emph{Social Science \& Medicine}}
  \bibinfo{volume}{191} (\bibinfo{year}{2017}), \bibinfo{pages}{168 -- 175}.
\newblock
\showISSN{0277-9536}
\urldef\tempurl%
\url{https://doi.org/10.1016/j.socscimed.2017.08.041}
\showDOI{\tempurl}


\bibitem[\protect\citeauthoryear{Twitter}{Twitter}{2019}]%
        {twitter}
\bibfield{author}{\bibinfo{person}{Twitter}.} \bibinfo{year}{2019}\natexlab{}.
\newblock \bibinfo{title}{Elections Integrity}.
\newblock
\newblock
\urldef\tempurl%
\url{https://about.twitter.com/en_us/values/elections-integrity.html}
\showURL{%
\tempurl}


\bibitem[\protect\citeauthoryear{Vincent}{Vincent}{2018}]%
        {Vincent}
\bibfield{author}{\bibinfo{person}{James Vincent}.}
  \bibinfo{year}{2018}\natexlab{}.
\newblock \bibinfo{title}{{This blessed Chrome extension replaces 'Elon Musk'
  with 'Grimes's Boyfriend'}}.
\newblock
\newblock
\urldef\tempurl%
\url{https://www.theverge.com/tldr/2018/5/10/17338984/elon-musk-grimes-boyfriend-chrome-extension}
\showURL{%
\tempurl}


\bibitem[\protect\citeauthoryear{Wang, Wang, and Zhu}{Wang
  et~al\mbox{.}}{2013}]%
        {Wang}
\bibfield{author}{\bibinfo{person}{Cheng-Jun Wang}, \bibinfo{person}{Pian-Pian
  Wang}, {and} \bibinfo{person}{Jonathan~J.H. Zhu}.}
  \bibinfo{year}{2013}\natexlab{}.
\newblock \showarticletitle{Discussing Occupy Wall Street on Twitter:
  Longitudinal Network Analysis of Equality, Emotion, and Stability of Public
  Discussion}.
\newblock \bibinfo{journal}{\emph{Cyberpsychology, Behavior, and Social
  Networking}} \bibinfo{volume}{16}, \bibinfo{number}{9}
  (\bibinfo{year}{2013}), \bibinfo{pages}{679--685}.
\newblock
\urldef\tempurl%
\url{https://doi.org/10.1089/cyber.2012.0409}
\showDOI{\tempurl}


\bibitem[\protect\citeauthoryear{Yang, Varol, Davis, Ferrara, Flammini, and
  Menczer}{Yang et~al\mbox{.}}{[n.d.]}]%
        {Cheng}
\bibfield{author}{\bibinfo{person}{Kai-Cheng Yang}, \bibinfo{person}{Onur
  Varol}, \bibinfo{person}{Clayton~A. Davis}, \bibinfo{person}{Emilio Ferrara},
  \bibinfo{person}{Alessandro Flammini}, {and} \bibinfo{person}{Filippo
  Menczer}.} \bibinfo{year}{[n.d.]}\natexlab{}.
\newblock \showarticletitle{Arming the public with artificial intelligence to
  counter social bots}.
\newblock \bibinfo{journal}{\emph{Human Behavior and Emerging Technologies}}
  \bibinfo{volume}{1}, \bibinfo{number}{1} (\bibinfo{year}{[n.\,d.]}),
  \bibinfo{pages}{48--61}.
\newblock
\urldef\tempurl%
\url{https://doi.org/10.1002/hbe2.115}
\showDOI{\tempurl}


\bibitem[\protect\citeauthoryear{{Yang}, {Wen}, {Lin}, and {Deng}}{{Yang}
  et~al\mbox{.}}{2017}]%
        {Yang}
\bibfield{author}{\bibinfo{person}{M. {Yang}}, \bibinfo{person}{X. {Wen}},
  \bibinfo{person}{Y. {Lin}}, {and} \bibinfo{person}{L. {Deng}}.}
  \bibinfo{year}{2017}\natexlab{}.
\newblock \showarticletitle{Quantifying Content Polarization on Twitter}. In
  \bibinfo{booktitle}{\emph{2017 IEEE 3rd International Conference on
  Collaboration and Internet Computing (CIC)}}. \bibinfo{pages}{299--308}.
\newblock
\urldef\tempurl%
\url{https://doi.org/10.1109/CIC.2017.00047}
\showDOI{\tempurl}


\bibitem[\protect\citeauthoryear{Zannettou, Caulfield, Setzer, Sirivianos,
  Stringhini, and Blackburn}{Zannettou et~al\mbox{.}}{2019}]%
        {Zannettou}
\bibfield{author}{\bibinfo{person}{Savvas Zannettou}, \bibinfo{person}{Tristan
  Caulfield}, \bibinfo{person}{William Setzer}, \bibinfo{person}{Michael
  Sirivianos}, \bibinfo{person}{Gianluca Stringhini}, {and}
  \bibinfo{person}{Jeremy Blackburn}.} \bibinfo{year}{2019}\natexlab{}.
\newblock \showarticletitle{Who Let The Trolls Out? Towards Understanding
  State-Sponsored Trolls}. In \bibinfo{booktitle}{\emph{Proceedings of the 10th
  ACM Conference on Web Science}} (Boston, Massachusetts, USA)
  \emph{(\bibinfo{series}{WebSci ’19})}. \bibinfo{publisher}{Association for
  Computing Machinery}, \bibinfo{address}{New York, NY, USA},
  \bibinfo{pages}{353–362}.
\newblock
\showISBNx{9781450362023}
\urldef\tempurl%
\url{https://doi.org/10.1145/3292522.3326016}
\showDOI{\tempurl}


\bibitem[\protect\citeauthoryear{Zuckerberg}{Zuckerberg}{2018}]%
        {fb}
\bibfield{author}{\bibinfo{person}{Mark Zuckerberg}.}
  \bibinfo{year}{2018}\natexlab{}.
\newblock \bibinfo{title}{Preparing for Elections}.
\newblock
\newblock
\urldef\tempurl%
\url{https://www.facebook.com/notes/mark-zuckerberg/preparing-for-elections/10156300047606634/}
\showURL{%
\tempurl}


\end{thebibliography}


\end{document}